\documentclass[2p,12pt,preprint,sort&compress]{elsarticle}

\usepackage{amsmath}
\usepackage{amssymb}
\usepackage{graphicx}
\usepackage{epsfig}
\usepackage{epstopdf}
\usepackage[margin=2.5cm]{geometry}
\usepackage{url}
\usepackage{hyperref}

\newcommand{\be}{\begin{equation}}

\newcommand{\ee}{\end{equation}}
\newcommand{\bea}{\begin{eqnarray}}
\newcommand{\eea}{\end{eqnarray}}
\newcommand{\bi}{\begin{itemize}}
\newcommand{\ei}{\end{itemize}}
\newcommand{\ben}{\begin{enumerate}}
\newcommand{\een}{\end{enumerate}}
\newcommand{\la}{\left\langle}
\newcommand{\ra}{\right\rangle}

\newcommand{\lp}{\left(}
\newcommand{\rp}{\right)}

\def\frac#1#2{{{#1}\over {#2}}}
\def\gsim{\mathrel{\rlap{\lower4pt\hbox{\hskip1pt$\sim$}}
    \raise1pt\hbox{$>$}}}         
\def\lsim{\mathrel{\rlap{\lower4pt\hbox{\hskip1pt$\sim$}}
    \raise1pt\hbox{$<$}}}         

\newcommand{\draft}[1]{}

\begin{document}


\title{Linear vs non-linear QCD evolution:\\ from HERA data to LHC phenomenology}


\author[label1,label2]{J.~L. Albacete}\ead{albacete@ipno.in2p3.fr}
\author[label3,label4]{J.~G. Milhano}\ead{guilherme.milhano@ist.utl.pt}
\author[label5]{P. Quiroga-Arias}\ead{pquiroga@lpthe.jussieu.fr}
\author[label4]{J. Rojo}\ead{juan.rojo@cern.ch}

\cortext[thanks]{Preprint number: CERN-PH-TH/2012-060}

\address[label1]{IPNO, Universit\'e Paris-Sud, CNRS/IN2P3, F-91406, Orsay, France.}
\address[label2]{IPhT, CEA/Saclay, 91191 Gif-sur-Yvette cedex, France.}
\address[label3]{CENTRA, Instituto Superior T\'ecnico, Universidade T\'ecnica de Lisboa, Av. Rovisco Pais, P-1049-001 Lisboa, Portugal.}
\address[label4]{Physics Department, Theory Unit, CERN, CH-1211 Gen\`eve 23, Switzerland.}
\address[label5]{LPTHE, UPMC Univ. Paris 6 and CNRS UMR7589, Paris, France}


\begin{abstract}
The very precise combined HERA data provides a testing ground in which the relevance of novel QCD regimes, other than the successful linear
DGLAP evolution, in small--$x$ inclusive DIS data can be ascertained. 
We present a study of the dependence of the AAMQS fits, based on the running coupling BK non-linear evolution equations (rcBK), on the fitted dataset.
This allows for the identification of the kinematical region where rcBK accurately describes the data, and thus for the determination of its applicability boundary.
We compare the rcBK results with NNLO DGLAP fits, obtained with the
NNPDF methodology with analogous kinematical cuts.
Further, we explore the impact on LHC phenomenology of
applying stringent kinematical cuts to the low--$x$ HERA
data in a DGLAP fit. 


\end{abstract}

\maketitle

\tableofcontents

\section{Introduction}
The knowledge of the partonic structure of the proton at all relevant observation scales plays a crucial role in the analysis of data from present high-energy hadronic colliders, most notably at the LHC. In practice such information is provided by phenomenological parton fits to previously existing data based on the use of perturbative QCD renormalization group equations and, in the framework of factorization theorems, it is used as input for establishing the theoretical expectations and uncertainties for the production rates of any process of interest.

The different QCD approaches for the description of the scale dependence of parton distribution functions -- analogously, for gauge invariant operators encoding the parton flux into the collision -- share a similar strategy in resumming to all orders radiative terms enhanced by large logarithms. 
The most widely used framework are the DGLAP equations~\cite{Dokshitzer:1977sg,Gribov:1972ri,Altarelli:1977zs}, that is, the renormalization group equations that describe the scale dependence of parton distribution functions through a resummation of large logarithms $\sim \alpha_{s}\ln Q^2/Q_0^2$ with $Q_{0}$ some initial scale. 
The DGLAP equations have been successfully and intensively tested against experimental data and, together with asymptotic freedom and factorization theorems,  provide a fundamental tool for establishing controlled predictions at the LHC. Successful as they are, the DGLAP equations are also expected to break down in some kinematic regimes. In particular, at small values of Bjorken-$x$, large logarithms $\sim\alpha_{s}\ln(x_{0}/x)$ emerge and need to be resummed to all orders. 



%

In turn, analogous resummation schemes aimed at describing the small-$x$ evolution of hadron structure have also been developed. In this direction in the kinematic plane, orthogonal to DGLAP evolution, the relevant logarithms are $\sim\alpha_{s}\ln(x_{0}/x)$, resummed to all orders in the BFKL approach~\cite{Kuraev:1977fs,Balitsky:1978ic}. Let us recall that although a consistent formulation of DGLAP evolution which also accounts for small-$x$ resummation has been formulated~\cite{Ciafaloni:2003rd,Altarelli:2008aj,Forte:2009wh,White:2006yh} in the recent years, its phenomenological consequences have not yet been fully explored. 
 Additionally, the enhancement of gluon emission at small-$x$ naturally leads to the -- empirically observed -- presence of large gluon densities and to the need of non-linear  recombination terms in order to stabilize the diffusion towards the infrared characteristic of BFKL evolution. Most importantly, the presence of non-linear terms is ultimately related to the preservation of unitarity of the theory.  

Both the resummation of small-$x$ logarithms and the inclusion of non-linear density dependent corrections are consistently accounted for by the B-JIMWLK~\cite{Balitsky:1996ub,Kovchegov:1999yj,Jalilian-Marian:1997gr,Kovner:2000pt,Weigert:2000gi} equations. The presence of non-linear terms in the small-$x$ evolution equations limits the growth rate of gluon number densities for modes of transverse momentum smaller than the saturation scale $Q_{s}$ (see section~\ref{bkvsdglap} for a precise definition). This novel, semi-hard dynamical scale marks the onset of non-linear corrections in QCD evolution and leads to distinctive dynamical effects such as the generation of geometric scaling~\cite{Stasto:2000er}. 

An important improvement in devising a phenomenological tool for the empiric study of the saturation phenomenon based on first principles QCD was brought by the recent determination of higher order corrections to the B-JIMWLK equations~\cite{Gardi:2006rp, Kovchegov:2006vj, Balitsky:2006wa, Balitsky:2008zza}. In particular, it was shown in \cite{Albacete:2007yr} that the BK equation~\cite{Balitsky:1996ub,Kovchegov:1999yj} -- the large-$N_{c}$ limit of the full B-JIMWLK hierarchy -- including running coupling corrections as calculated in ~\cite{Kovchegov:2006vj,Balitsky:2006wa,Gardi:2006rp} was compatible with experimental data from different collision systems. 
This expectation was confirmed by the AAMQS collaboration in~\cite{Albacete:2009fh}, where global fits to e+p data based on the use of the  running coupling BK equation (referred to as rcBK equation henceforth) were performed for the first time, providing a good description of data. This result was then updated in~\cite{Albacete:2010sy} to include in the fitted data set the combined H1+ZEUS data~\cite{:2009wt} on the reduced cross section and also include the contribution of heavy quarks. The AAMQS data set covers the region $x<0.01$ and $Q^{2}<50$ GeV$^{2}$, including data in the photoproduction region where $Q^{2}\ll1$ GeV$^{2}$. A similar strategy was followed in~\cite{Kuokkanen:2011je}, where data on the diffractive cross section was also fitted consistently. Overall, a very good description of data is provided in those three studies. Although the rcBK fitting technology has not yet reached the level of sophistication and accuracy of DGLAP based approaches, it shares a similar strategy and methodology, thus allowing for systematic comparison to DGLAP studies, as we intend to do in this work.

Theoretical arguments alone can only strictly establish the applicability of either DGLAP or BK in their asymptotic limits of very large $Q^2$ (for DGLAP) or very small $x$ (for BK).  
On the phenomenological side, where intermediate $(x,Q^{2})$ kinematics is probed, the situation remains unclear. On one hand, DGLAP based approaches have continuously reported good fits to all available data above some initial scale $Q_{0}^{2}\sim 1\div 4$ GeV$^{2}$ and for $x$-values as small as $\sim10^{-5}$. On the other hand, similarly good fits to data in the same kinematic region are achieved using the rcBK evolution equation, including data points up to $Q^{2}=50$ GeV$^{2}$. Moreover, some features identified in data such as geometric scaling or the ratio of total over diffractive cross sections are naturally explained in a non-linear evolution framework, while they are difficult to reconcile with the strictly linear approach. Although geometric scaling has been shown to be fully compatible with the linear DGLAP equations~\cite{Caola:2008xr}, the dynamical scale generation at its origin in the non-linear context is not viable in a linear framework.
 
Clearly, the reliability of one or the other approach in the region of moderate $(x,Q^2)$ cannot be determined on a priori theoretical arguments. It is also clear that claims in favor of one particular approach should not be done solely on the basis of agreement with experimental data: it is well known that
one can obtain an excellent fit to the HERA low--$x$ data with
a very reduced number of free parameters~\cite{Forte:1995vs}. Such is a necessary but not sufficient condition.  Further, beyond describing existing data, the usefulness of a given approach rests on its predictive power towards kinematic regions experimentally unexplored so far. This latter condition rules out phenomenological data descriptions such as the structure function parametrizations
of Refs.~\cite{Lastovicka:2002hw,DelDebbio:2004qj}. The predictivity requirement is clearly satisfied by both the BK and DGLAP approaches, both endowed with a well defined QCD dynamical input that we briefly review in Sect.~\ref{bkvsdglap}. Their predictive power is, however, oriented towards different directions in the kinematic plane: small-$x$ and high-$Q^{2}$ respectively.

A pertinent question to ask is whether corrections to the limit in which both formalisms are well defined, large-$Q^{2}$ and small-$x$ respectively, are important in intermediate kinematic regions and, if so, to what extent this missing dynamical effects could be accommodated in the boundary conditions for the respective rcBK or DGLAP evolution. 
Thus, it may be that the flexibility in the initial conditions for DGLAP evolution is hiding some interesting QCD dynamics, namely the presence of non-linear behavior or the onset of the regime dominated by high energy QCD (small--$x$ resummation) not included in DGLAP fits at their present degree of accuracy. A first attempt to answer this question was carried out in \cite{Caola:2009iy,Caola:2010cy}.  Also, it is conceivable that higher twist corrections not included in the collinear factorization framework may be sizable at moderate and small $Q^{2}$. Analogous concerns arise in relation to the applicability of the rcBK approach as it is conceivable that the maximal value $x_{0}=0.01$ considered in the AAMQS fits might not be small enough for the dipole formalism they rely upon to be fully applicable. 
In particular, it remains to be
understood why data at $Q^2=50$ GeV$^2$ can be described in a rcBK fit, far beyond the naive range of application of the nonlinear description. 
In that case, sizable energy conservation corrections, negligible at very small-$x$, may influence the values of the fit parameters.

A natural step towards elucidating whether interesting dynamics is hidden or absorbed in the boundary conditions in either approach is to systematically displace those boundaries and assess the fit stability under such changes or, alternatively, look for correlations between fit parameters and the position of the boundary. These boundary conditions are the Parton Distribution Functions (PDFs)
for DGLAP evolution and the Unintegrated Gluon Distribution (UGD) for
BK evolution.
%
%
Such sensitivity would indicate that the resulting PDFs (in the case of DGLAP)  or UGDs (in the case of rcBK) extracted from the fit are contaminated with physics effects beyond the dynamical content of the respective evolution equations.
%
%
In Sect.~\ref{FS} we describe the strategy pursued in \cite{Caola:2009iy,Caola:2010cy} to search for systematic deviations of DGLAP fits by imposing selected kinematic cuts to the data set. We shall extend it in an analogous fashion to the framework of rcBK fits. The results for the updated DGLAP and rcBK fits to HERA data with kinematic cuts are presented and discussed in Sects.~\ref{nnlocuts} and \ref{cutrcBK} respectively.

The presence of any systematic deviations from fixed order DGLAP evolution will have direct implications for LHC phenomenology: indeed,
parton distributions extracted from a DGLAP analysis which includes HERA 
data in the small-$(x,Q^{2})$ region are then evolved upwards in $Q^2$ to predict any LHC processes. In Sect.~\ref{implications} we shall quantify and explore how this potential source of theoretical uncertainty propagates towards high $Q^{2}$ scales by computing benchmark LHC cross sections with PDF sets both with and without the small--$x$ kinematical cuts.
Conclusions and outlook are presented in Sect.~\ref{conclusions}.


\section{rcBK and DGLAP evolution}
\label{bkvsdglap}

In order to frame  the discussion and facilitate the interpretation of the results presented in the next sections, and before describing the strategy adopted to chart the small-$(x,Q^{2})$ kinematic territory, we first review very succinctly the main features of both BK and DGLAP approaches.
%
Using rather compact notation\footnote{The BK equation written here assumes translational invariance of the dipole amplitude $\mathcal{N}$. Also, some constant factors have been absorbed in the evolution kernel $\mathcal{K}$, see e.g.~\cite{Albacete:2010sy} for precise definitions.} the BK (for the dipole  scattering amplitude $\mathcal{N}(x,r)$) and DGLAP evolution equations (for vector PDFs $f(x,Q^{2}$)) can be written as follows:
\begin{equation}
\text{BK:}\quad\frac{\partial\mathcal{N}(r,x)}{\partial\ln(x_{0}/x)}\!=\!\int d^{2}r_{1}\mathcal{K}(r,r_{1},r_{2})\left[\mathcal{N}(r_{1},x)\!+\!\mathcal{N}(r_{2},x)\!-\!\mathcal{N}(r,x)\!-\!\mathcal{N}(r_{1},x)\mathcal{N}(r_{2},x)\right]\,; 
\label{bkeq}
\end{equation}

\begin{equation}
\text{DGLAP:}\quad{\partial f(x,Q^2) \over \partial \ln (Q^2/Q_{0}^{2})}=\int_{x}^{1}\frac{dy}{y}   P\left(\alpha_s(Q^2),x/y\right)  f(y,Q^2)\,.
\label{dglapeq}
\end{equation}

The object evolved by the BK equation is the (imaginary part of the) quark-antiquark color dipole amplitude to scatter off a hadronic target, $\mathcal{N}(r,x)$, where $r$ is the transverse coordinate and $x$ is the usual Bjorken scaling variable in deep--inelastic scattering. The onset of the black-disk limit is given by the condition ${\mathcal{N}}(r_s=1/Q_s(x),x)=\kappa\sim 1$, which can also be used to define the saturation scale, $Q_s(x)$.
The amplitude  $\mathcal{N}(r,x)$  is the main dynamical ingredient of the dipole formulation of deep inelastic scattering, extensively used in the phenomenological searches of non-linear dynamics in data, including the AAMQS global fits. In this framework the virtual photon-proton cross section for transverse ($T$) and longitudinal ($L$) polarization of the virtual photon reads

\begin{equation}
  \sigma_{T,L}(x,Q^2)=2\sigma_{0}\sum_f\int_0^1 dz\int d{\bf r}\,\vert
  \Psi_{T,L}^f(e_f,m_f,z,Q^2,{\bf r})\vert^2\,
  {\cal N}(r,x)\,,
\label{dm}
\end{equation}
where $\Psi_{T,L}^f$ is the light-cone wave function for a virtual photon to fluctuate into a quark-antiquark dipole of quark flavor $f$. Other observables of interest, such as structure functions $F_{2}$, $F_{L}$, are straightforwardly related to $\sigma_{T,L}$ in Eq.~(\ref{dm}). A clearer physical interpretation of the dipole amplitude is obtained by recalling that its Fourier transform yields the unintegrated gluon distribution (UGD) of the target:  $\phi(x,k_{t})=\int d^{2}r\,e^{i\vec{r}\cdot\vec{k_t}}\mathcal{N}(r,x)$. In turn, to LO accuracy the UGD is related to the standard integrated gluon distribution: $xg(x,Q^2)=\int^{Q^2}\!d^2 k_t\, \phi(x,k_T)$. Thus, the dipole amplitude provides direct information on the gluon content of the hadron.
The kernel $\mathcal{K}$ of the BK equation is obtained via perturbative QCD calculations for small-$x$ gluon emission, in a fashion analogous to that employed for the calculation of the DGLAP splitting functions, $P\lp x,\alpha_s\rp$ in Eq.~(\ref{dglapeq}). The evolution kernel has been calculated up to NLO accuracy in the resummation variable $\alpha_{s}\ln (x_{0}/x)$~\cite{Balitsky:2008zza}. Alternatively, a partial subset of NLO corrections can be resummed to all orders to obtain the running coupling BK equation used later in this work.


Three main distinctive features of the BK equation deserve to be highlighted. First, it is an evolution equation in Bjorken-$x$, i.e. it provides the change of hadron structure when smaller values of $x$ are probed.  As such, it has no predictive power in the orthogonal $Q^{2}$-direction, with all the $Q^2$ dependence of structure functions computed with rcBK fixed by the wave function in Eq.~(\ref{dm}).
Second, it is a non-linear equation. The presence of non-linear terms is required by preservation of unitarity\footnote{It is indeed straightforward to check that the non-linear term in Eq.~(\ref{bkeq}) prevents the dipole scattering amplitude to grow above unity, provided the initial condition is unitary itself, i.e $\mathcal{N}\leq 1$.}, and can be interpreted, in the appropriate gauge and frame, as due to gluon recombination processes. 
Third, it resums all twists or multiple scatterings, and thus is applicable also for small values of $Q^{2}$.

In contrast, the DGLAP equation, Eq.~(\ref{dglapeq}), provides the $Q^{2}$ evolution of proton structure. The matrix of splitting functions that govern the evolution are known
up to $\mathcal{O}\lp\alpha_s^2\rp$ in pQCD~\cite{gNNLOa,gNNLOb}. 
It has no predictive power in the orthogonal $x$-direction: in a DGLAP analysis, for values
of $x\le x_{\rm min}$, with $x_{\rm min}$ the smallest value of $x$ of the data
included, the DGLAP predictions become unreliable.  Also, it is a linear, leading twist equation, two conditions that are expected to break down for sufficiently small values of $Q^{2}$, where gluon densities are higher and the contribution from higher twists may be important.

Contact between these two orthogonal approaches can be made in the diagonal limit of $x\to0$ and $Q^{2}\to\infty$, where both BK and DGLAP equations converge to the Double Logarithmic Approximation (see e.g.~\cite{Kovchegov:1999yj}). However, no smooth interpolation between the two is known to date in the more interesting phenomenological region of moderate $x$ and $Q^{2}$. 


Both the DGLAP or the BK equations are initial value problems, i.e. they are well defined only after initial conditions at the initial evolution scale have been provided. 
%
%
This introduces free parameters, ultimately of non-perturbative origin, to be fitted to data.
In the case of BK, initial conditions for the dipole amplitude (the UGD) should be specified at the initial evolution scale $x=x_{0}$ and for all values of the the dipole size $r$ (equivalently, for all values of\footnote{The  photon wave function $\Psi_{T,L}^f$ in Eq.~(\ref{dm}) peaks at $r\sim2/Q$.} $Q^{2})$. In the AAMQS rcBK fits to HERA data the initial conditions are taken in the form 
\begin{equation}
\mathcal{N}(r,x=x_{0})=1-\exp\left(-\frac{(r^{2} Q_{s0}^{2})^{\gamma}}{4}\ln\left(\frac{1}{r\Lambda_{\rm QCD}}+e\right)\right) \,,
\end{equation}
where $Q_{s0}$ and $\gamma$ are free parameters to be determined
from the data. For DGLAP, initial conditions for the PDFs have to be provided at some low initial scale $Q_{0}^{2}$ and for all values of $x$: $xf(Q^{2}=Q^{2}_{0},x)$. Within the NNPDF
approach, these initial conditions are parametrized with artificial neural networks, which provide universal interpolants and avoid any theoretical bias
arising from the choice of a particular functional form for the input PDFs.
Note that, in principle, this technology could also be applied for
the initial conditions of the rcBK analysis.

Lacking any better quantitative criterion, the choice of initial scale -- $x_{0}$ for BK or $Q_{0}^{2}$ for DGLAP -- that limits the fitting the data set in global fits is commonly determined a posteriori: if good quality fits to data can be obtained with a given choice of the initial scale, then the validity of the applied evolution scheme down to such scale is accepted.
However, this empirical, self-consistent criterion can be a misleading one. It is perfectly conceivable that the adjustable parameters entering the initial conditions in either evolution scheme may absorb some relevant dynamics not properly described by the physics content comprised in the equations themselves. 


\label{FS}
Thus, one needs to define some suitable strategy to identify the regimes
of validity of each formalism and quantify the potential deviation
from these.
Such a strategy to search for statistically significant deviations from DGLAP evolution was laid down in Refs.~\cite{Caola:2009iy,Caola:2010cy}. There, subsets of data on the reduced lepton--proton DIS cross section $\sigma_{r}(x,Q^{2})$ measured at HERA~\cite{:2009wt}, below some given kinematic cuts,
\be
 Q^{2} \le Q^{2}_{\rm cut} \equiv  A_{\rm cut}\,x^{-\lambda}
\label{eq:cut}
\ee
 with $\lambda \sim 0.3$ and different values of $A_{\rm cut}$, were excluded from the data set, so that the PDFs were fitted excluding the potentially dangerous region. The cuts were motivated by the generic expectation that possible 
deviations from fixed order DGLAP are larger at small--$x$ and $Q^2$. 
Then, backwards DGLAP evolution was used to compare with the excluded data, thus providing a direct test of fit stability under changes in the boundary conditions. 

The analysis of~\cite{Caola:2009iy,Caola:2010cy}, carried at NLO
in the massless scheme for structure functions, 
found a systematic discrepancy, albeit with not large enough statistical significance for a decisive statement to be made.
The observed discrepancy indicated that the parton evolution in the unfitted kinematic region may not be fully accounted for by the physics encoded in NLO DGLAP equations. Rather, additional dynamics should be responsible for it. 
Such deviations are qualitatively consistent with the behavior predicted by small-$x$ perturbative resummation~\cite{Altarelli:2008aj}, but incompatible with next-to-next-to-leading order corrections.  Also, it was suggested that an improved treatment of the heavy quark masses may have a sizable impact for the relatively low $Q^{2}$ values in the region excluded by the saturation inspired
kinematical cuts mentioned above.

\begin{figure}[t]
\begin{center}
\includegraphics[height=12cm]{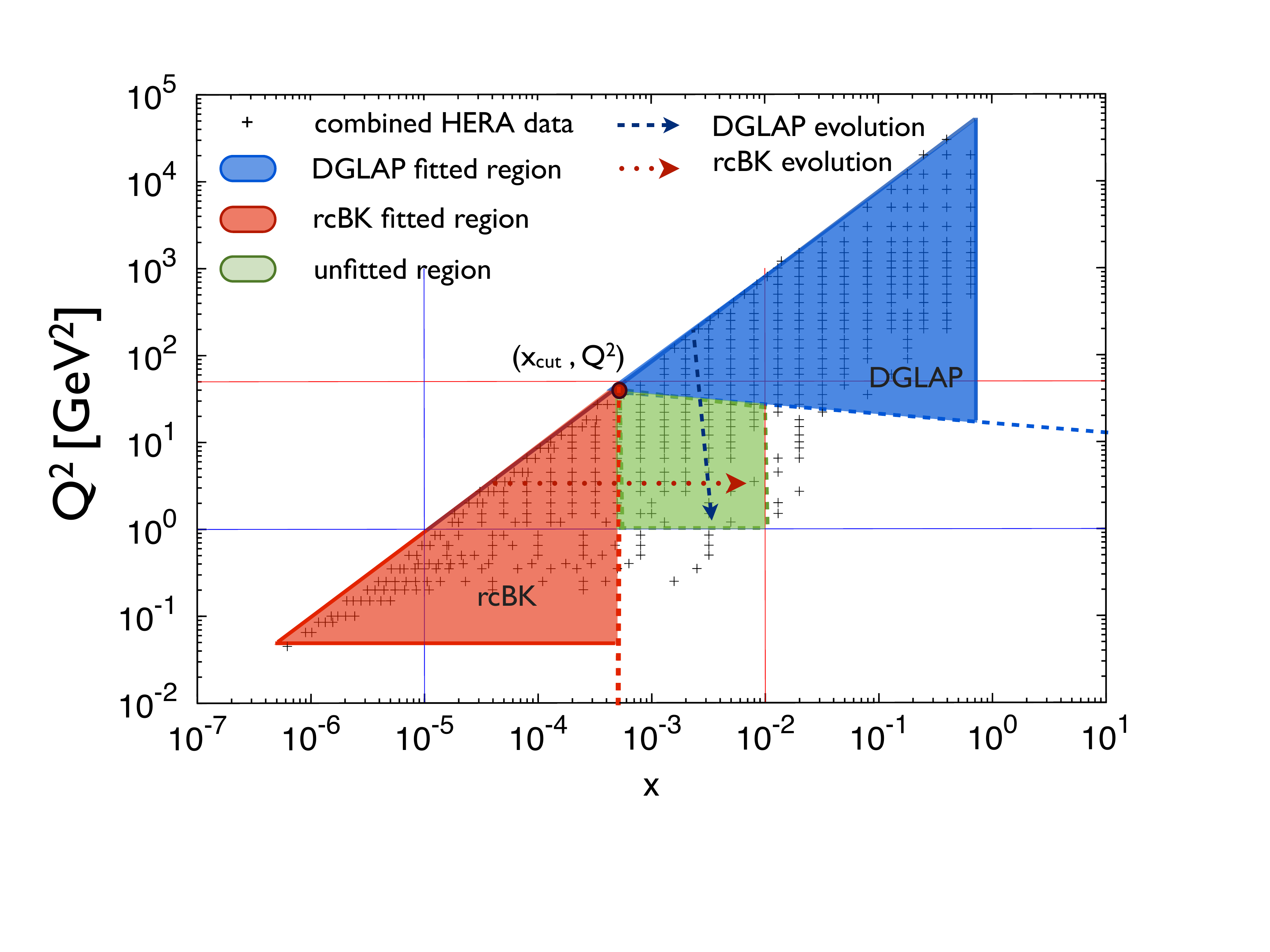}
\end{center}
\vspace{-2cm}
\caption{\small Sketch of the kinematic plane with cuts for DGLAP and rcBK fits. The arrows indicate backwards evolution in either formalism to the unfitted {\it test} region. }
\label{method}
\end{figure}

In this work, we shall study the stability of the AAMQS fits with respect to
the choice of dataset following an analogous procedure. 
The AAMQS fits to HERA data, based on the rcBK non--linear
evolution equations 
included data in the region $x<x_{0}=0.01$ and $Q^{2}<50$ GeV$^{2}$.
We shall perform fits to data using the AAMQS set up for the case of only three active flavors (the lightest ones) with 4 free parameters ($Q_0^2$, $\gamma$, $\sigma_0$ and $C$) as described in detail in~\cite{Albacete:2010sy}. We shall systematically reduce the largest experimental value of $x$ included in the fit, which we denote by $x_{\rm cut}$, and then use the parametrization for the dipole scattering amplitude resulting from the fit to predict the value of the reduced cross section in the unfitted region $x_{\rm cut}<x<x_{0}$. 

The rcBK equation at the basis of the AAMQS approach is a non-linear equation and, therefore, extremely unstable under backwards evolution. Equivalently, as it is well known, the solutions of the BK equation at asymptotically small-$x$ are universal, i.e independent of the initial conditions. This immediately implies that backward BK evolution is not well defined. Thus, even for fits with a cut in the Bjorken variable at $x_{\rm cut}$, we shall start the evolution, and hence determine the boundary conditions, at $0.01=x_{0}>x_{\rm cut}$. This will allow us to know the dipole scattering amplitude corresponding to fits with cuts in all the unfitted region, i.e. $x_{\rm cut}<x<x_{0}$. The fitting strategy 
for the DGLAP and rcBK analysis with kinematical cuts 
is summarized in Fig.~\ref{method}.


\section{NNLO DGLAP analysis with kinematical cuts}
\label{nnlocuts}

Before presenting the results for the rcBK fits with cuts,
we present an updated DGLAP fit with kinematic cuts, along
the lines of those presented in  Refs.~\cite{Caola:2009iy,Caola:2010cy},
but now based on the NNPDF2.1 NNLO set~\cite{Ball:2011uy}.
NNPDF2.1 NNLO is based on the FONLL-C GM-VFN scheme~\cite{Forte:2010ta}
for an accurate
treatment of heavy quarks in DIS, while the original 
studies~\cite{Caola:2009iy,Caola:2010cy} 
where based on PDFs sets extracted in the ZM-VFN 
scheme~\cite{Ball:2008by,Ball:2009mk,Ball:2010de}. 
Although we will refrain from a detailed discussion of
the NNPDF2.1 NNLO cut fits that were produced for comparison
with the rcBK fit with cuts  presented below, we briefly explain our modus operandi and highlight some
of its features. 

\begin{figure}[h]
\centering
\epsfig{width=0.49\textwidth,figure=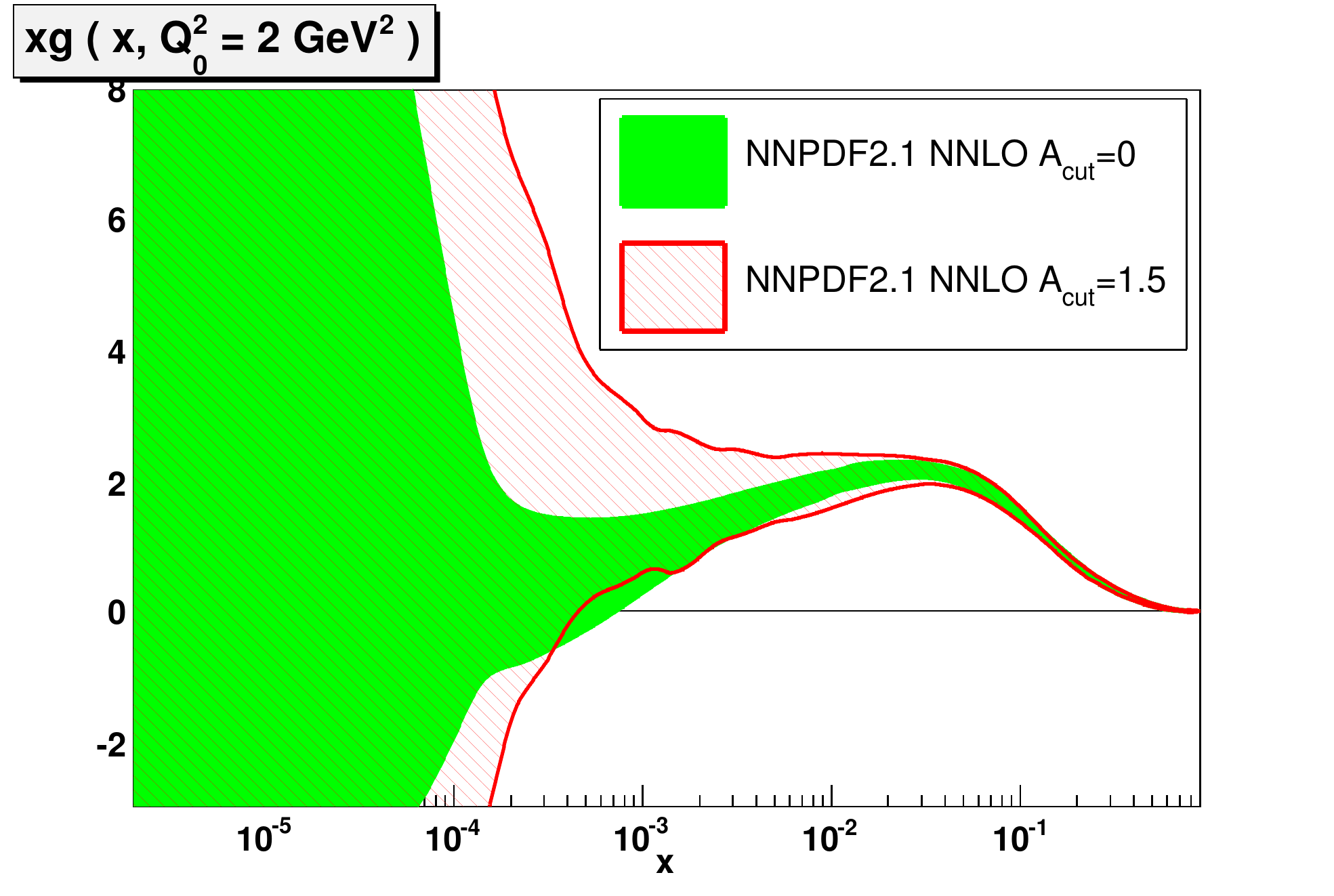}
\epsfig{width=0.49\textwidth,figure=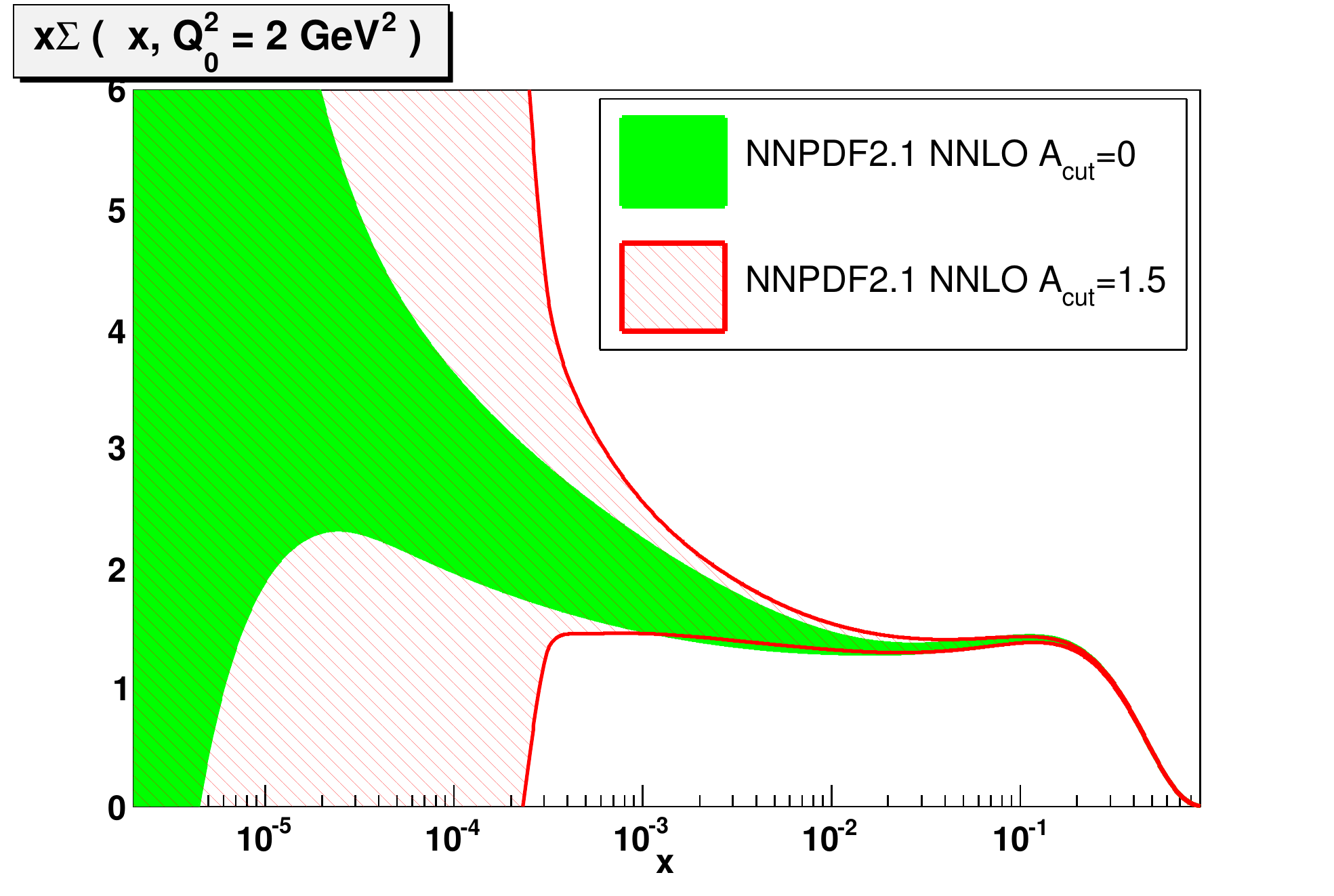}
\epsfig{width=0.49\textwidth,figure=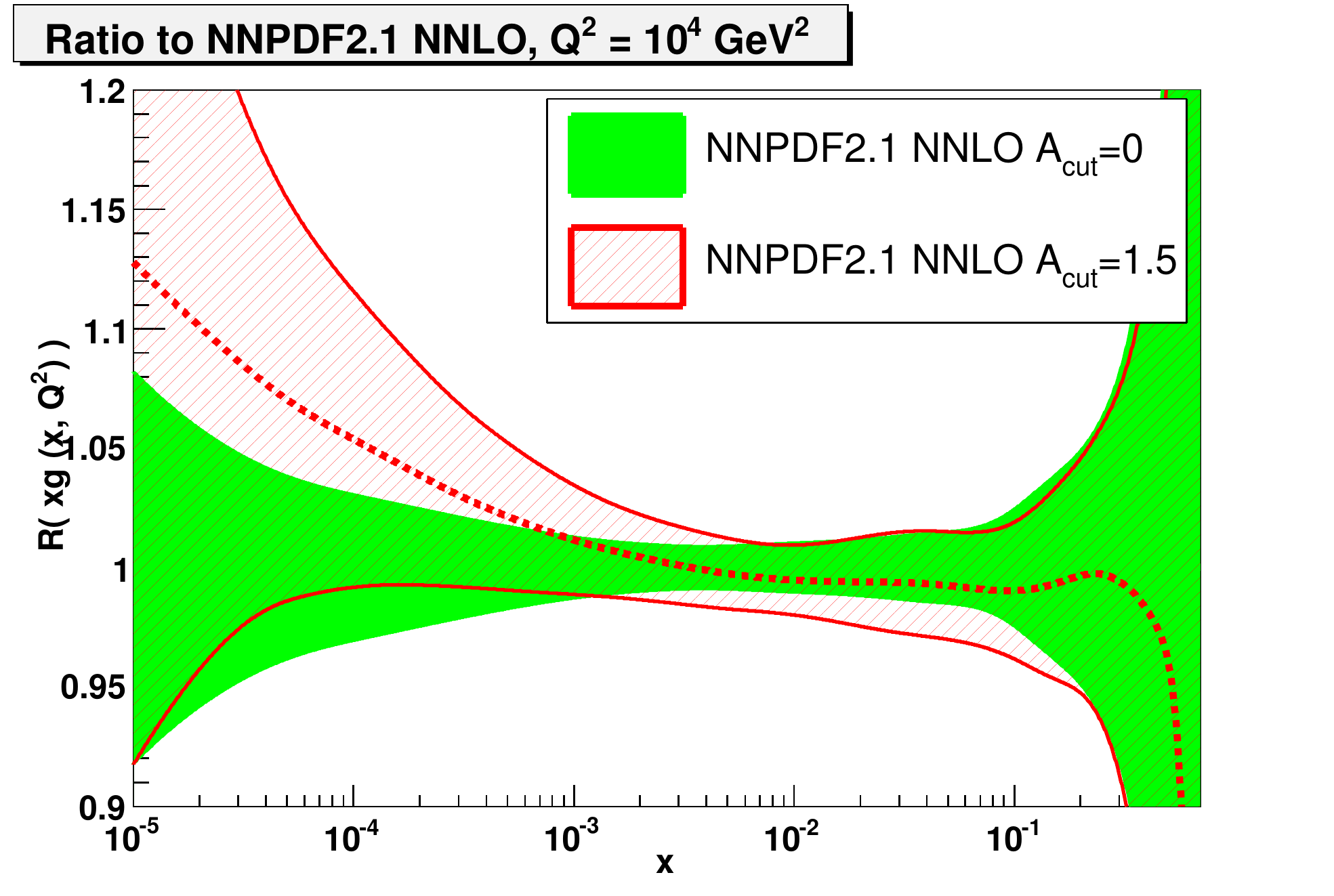}
\epsfig{width=0.49\textwidth,figure=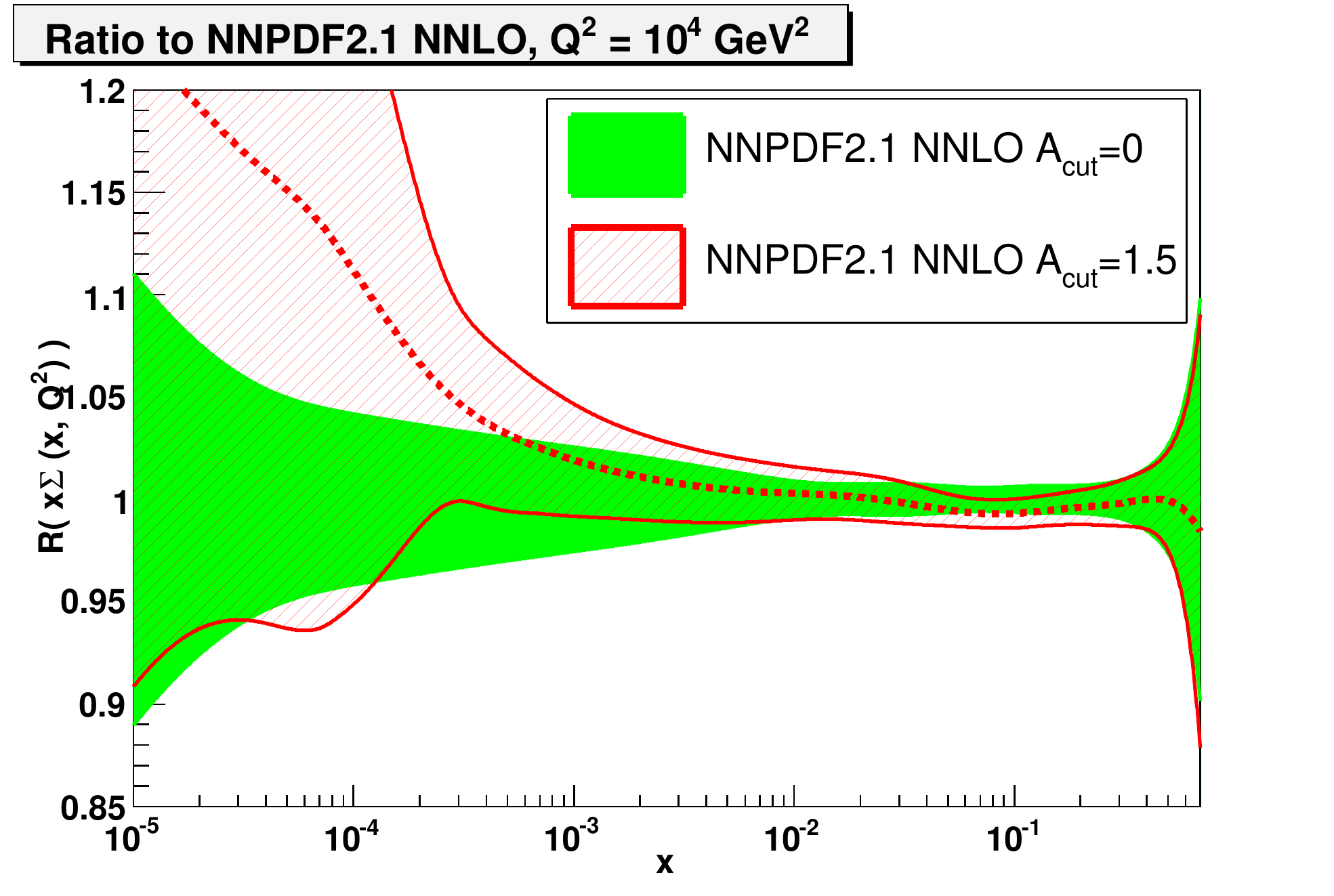}
\caption{\small Comparison between the NNPDF2.1 NNLO
PDFs with $A_{\rm cut}=1.5$ and $A_{\rm cut}=0$ for the gluon
(left plots) and the quark singlet (right plots). We perform
the comparison of the absolute PDFs at the initial scale
$Q^2_0=2$ GeV$^2$
in the upper plots, and we compare the ratio with respect the
reference set at a typical LHC scale $Q^2=10^4$ GeV$^2$
in the lower plots.
\label{fig:pdfs}}
\end{figure}

The reference DGLAP fit is NNPDF2.1, a NNLO fit to all data ($A_{\rm{cut}}=0$) above $Q^2>3$ GeV$^2$, including ${\cal{O}}$(3500) experimental points from DIS, DY, $W/Z$ production and inclusive jet data. 
Then, based on that fit,  the $\chi^2$ per number of data points\footnote{In the NNPDF analysis the full information on the correlated systematics
is taken into account into the definition of the $\chi^2$, 
and the normalization uncertainties are included following
the $t_0$ prescription~\cite{Ball:2009qv}. Let us recall
that non negligible differences are expected if the systematic
uncertainties are added in quadrature to the statistical error,
and that the $\chi^2$ will be artificially about 15\% lower in this
latter approximation~\cite{:2009wt}.}  is calculated for different subsets of interest.  Of relevance to this work is the  neutral current positron subset, HERA-I NCp, since these are the only data included in the rcBK analysis, and also the only data affected by the kinematic cuts. We have then performed a new NNLO DGLAP fit including only those data points which survive the cut $A_{\rm{cut}}=1.5$.  
The NNLO PDFs with $A_{\rm cut}=1.5$, see Eq.~(\ref{eq:cut}), are
compared to the reference NNPDF2.1 NNLO PDFs (with $A_{\rm cut}=0$)
in Fig.~\ref{fig:pdfs}, where the two
PDFs that are mostly affected by the applied kinematical cuts -- the singlet and the gluon -- are shown,
both at the initial scale and at  $Q^2=10^4$ GeV$^2$, a typical value for LHC phenomenology.
In the latter case we show the ratios with respect to the reference
uncut fit. 
This comparison shows that while PDF uncertainties increase
substantially when the low--$x$, low--$Q^2$ HERA data are removed, 
the two PDF sets are always consistent at the one-sigma level. The
differences remain when DGLAP evolution is used to determine the
PDFs at a typical LHC scale of  $Q^2=10^4$ GeV$^2$ where both
the gluon and the singlet are larger at small-$x$ as compared
to the uncut fit.



\begin{figure}[h]
\centering
\epsfig{width=0.49\textwidth,figure=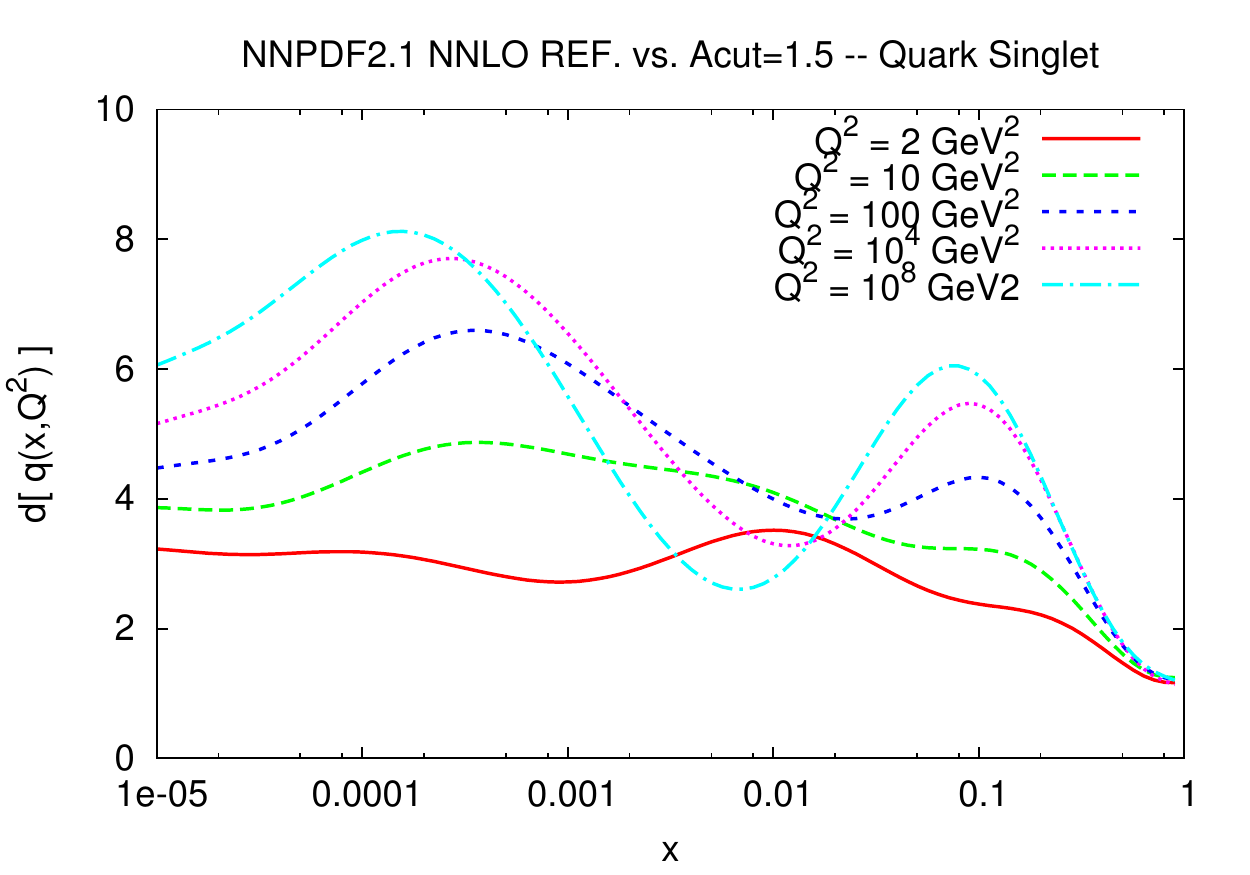}
\epsfig{width=0.49\textwidth,figure=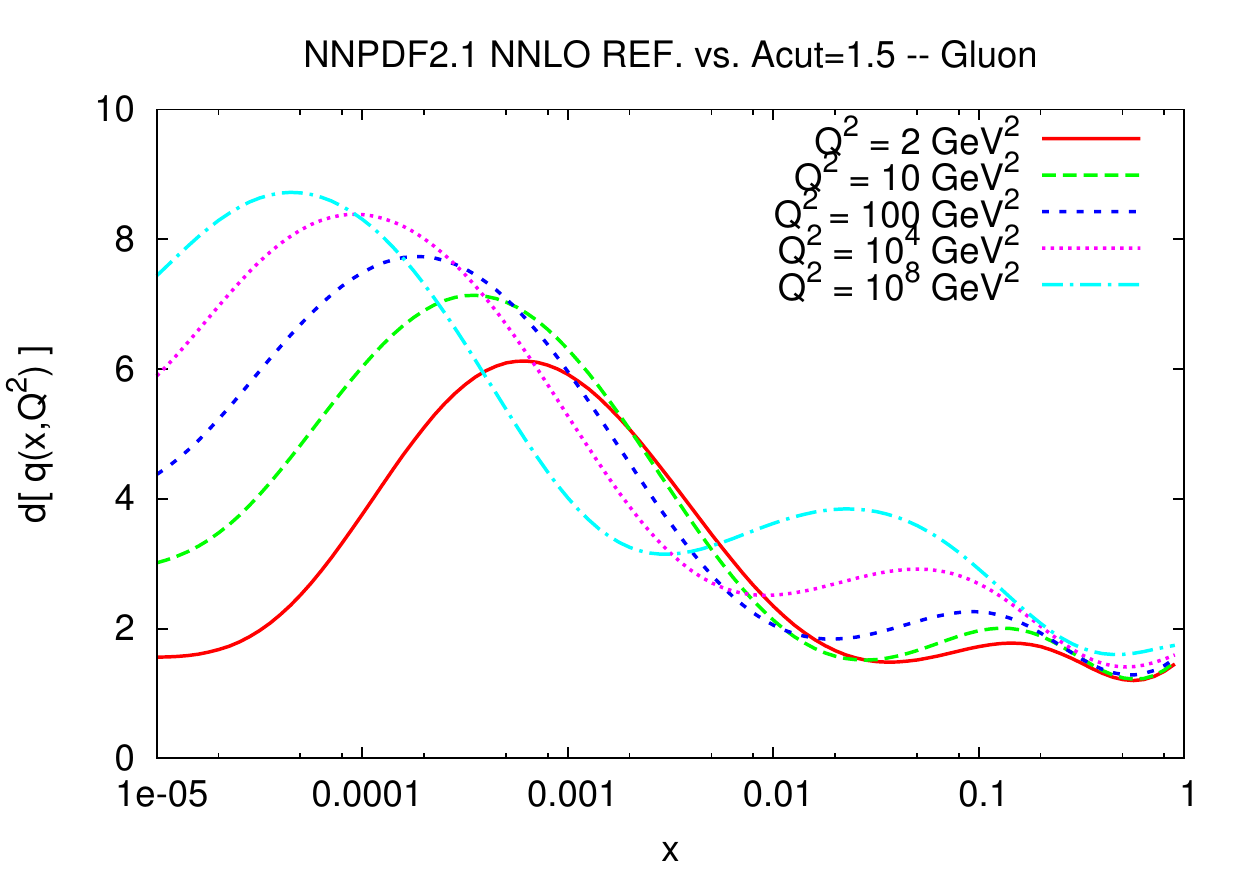}
\caption{\small Distances between PDF central values as a function
of the scale $Q^2$ of the PDFs. We show the results for the quark
singlet PDF (left plot) and for the gluon PDF (right plot).
\label{fig:distances}}
\end{figure}

It is known that in the small--$x$ and large $Q^2$ limit,
the trajectories dictated by DGLAP evolution become independent
of the initial condition. To quantify if this asymptotic
regime is reached for realistic kinematics, we show
 the distances between PDFs\footnote{The distances
defined to compare PDFs in the NNPDF framework
are defined in the Appendix B of Ref.~\cite{Ball:2010de}.
They should not be confused with the relative and statistical
distances, eqs.~(\ref{eq:drel}) and~(\ref{eq:dstat}),  introduced later in this work.
} as the 
scale is raised in 
Fig.~\ref{fig:distances}. 
This comparison allows us to
quantify how DGLAP evolution modifies the differences
in the initial scale PDFs as the scale is raised.
 Let us recall that once the
initial conditions are fixed at some low scale $Q_0^2$, the
PDFs at any other scale $Q^2\ge Q_0^2 $ are completely determined
by the DGLAP evolution equations. As one can see, DGLAP
evolution modifies the PDFs but does not wash up the initial differences,
even for scales as large as $Q^2=10^8$ GeV$^2$, shifting these differences towards smaller--$x$ values, as clearly
seen for example in the gluon PDF.


Other PDF fitting groups have investigated the same topic, with
inconclusive results so far. In the CT collaboration study~\cite{Lai:2010vv}, based on the CT10 NLO analysis, the dependence of the fit
quality to the HERA--I data with the same kinematical cuts as
in Refs.~\cite{Caola:2009iy,Caola:2010cy} was studied. 
It was found that some deviation for the smallest $x$
and $Q^2$ was obtained only for given choices of PDF parametrizations,
while for other parametrizations the results were consistent
with NLO QCD. This is not in contradiction with the NNPDF
analysis, which instead probe simultaneously a wide range
of possible initial conditions. 
A full clarification would require, within the
CT framework, the exploration of a larger variety
of initial PDF conditions in order to statistically quantify 
the significance of the possible deviations with respect to fixed order
DGLAP evolution.

The HERAPDF collaboration has also studied the impact of
raising the kinematical cuts and assessed the variation of fit quality when
going from NLO to NNLO. In Ref.~\cite{H1prelim-10-044}
 a NLO QCD analysis was performed
on the combined HERA-I data supplemented by HERA--II taken
at lower beam energies. When the kinematical cut was raised
to $Q^2\ge 5$ GeV$^2$ the resulting small-$x$ singlet and
gluon were found to be outside the reference PDF
uncertainty band. Also, HERAPDF1.0 NNLO fits lead to a worse
fit quality (by more that 50 units in $\chi^2$ for about 500
data points) than the corresponding NLO results. Similar
conclusions are found in the updated HERAPDF1.5 NLO and
NNLO analysis~\cite{H1prelim-11-042}. There
a larger PDF uncertainty on the gluon distribution is found at NNLO
as compared to NLO. 

\begin{figure}[t]
\begin{center}
\vspace{-0.5cm}
\includegraphics[height=10cm]{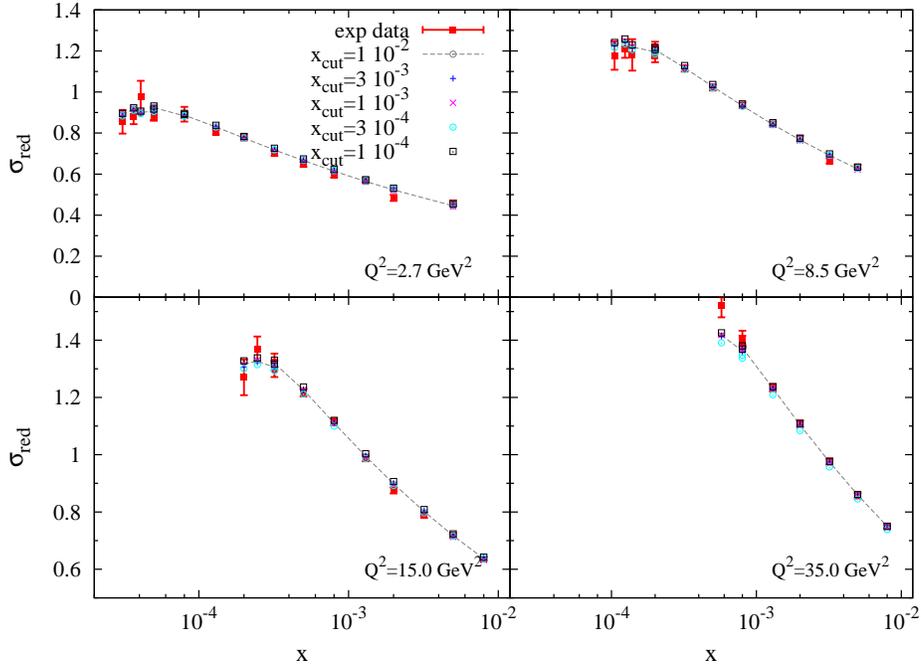}
\end{center}
\vspace{-0.5cm}
\caption{Comparison of the result for reduced cross section obtained with rcBK fits with different cuts: $x_{\rm cut}=10^{-2},\, 3\cdot 10^{-3},\,10^{-3},\, 3\cdot10^{-4}$ and $10^{-4}$ and HERA data for four different bins in $Q^{2}=2.7,\,8.5,\,15$ and 35 GeV$^{2}$.}
\label{rcBKdata}
\end{figure}

\section{rcBK analysis with kinematical cuts}
\label{cutrcBK}

We now present the rcBK results with various kinematical cuts
as described in Sect.~\ref{bkvsdglap}, and compare them with the
DGLAP cut fits presented in Sect.~\ref{nnlocuts}.
Fig.~\ref{rcBKdata} shows the comparison of the theoretical results stemming from fits to data with different $x$-cuts ($x_{\rm cut}=10^{-2},\, 3\cdot 10^{-3},\,10^{-3},\, 3\cdot10^{-4}$ and $10^{-4}$) to experimental data  on the reduced cross section for $x<10^{-2}$ and $Q^2<50$ GeV$^2$. Several comments are in order. First, fits to data with different cuts yield comparably good 
$\chi^{2}/{\rm d.o.f.}$\footnote{In the AAMQS approach the $\chi^{2}$ is calculated as $\chi^{2}=\sum_{i}\frac{(\sigma_{\rm r,th}-\sigma_{\rm r,exp})^{2}}{\Delta\sigma_{\rm r,exp}^{2}}$, where $\Delta\sigma_{\rm r,exp}^{2}$ is the total experimental error obtained adding in quadrature
all experimental uncertainties, and thus neglects the effects of correlations.}, despite the decreasing number of points with decreasing $x_{\rm cut}$.  Also, the extrapolations of the results for the reduced cross section from fits with cuts to the unfitted region , i.e to $x>x_{\rm cut}$, yield a good description of the data. 
This illustrates the stability of the AAMQS fits under changes in the boundary conditions and lends support to the idea that the non-linear small-$x$ dynamics comprised in the running coupling BK equation describes well the scale dependence of the proton structure in this test region of moderate values of $x\lesssim0.01$ and $Q^{2}\lesssim 50$ GeV$^2$. This result is in contrast with the DGLAP cut fits presented in the previous section, where systematic deviations among fits with and without cuts were observed.

\begin{figure}[t]
\begin{center}
\vspace{-0.5cm}
\includegraphics[height=10cm]{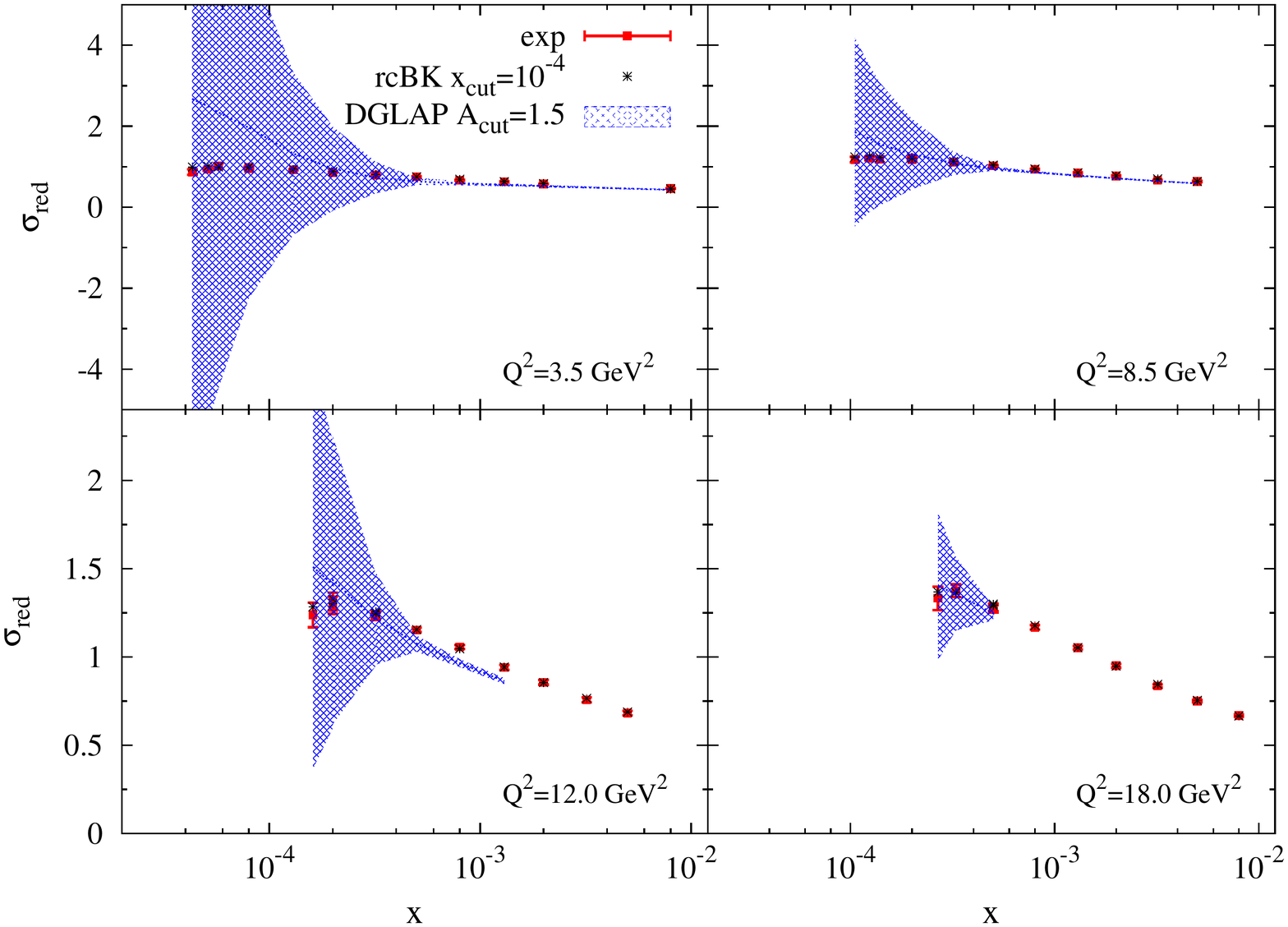}
\end{center}
\vspace{-0.5cm}
\caption{\small Reduced cross section obtained with the rcBK cut fit with $x_{\rm cut}=10^{-4}$ and the DGLAP fit with $A_{\rm cut}=1.5$, compared to
the experimental HERA-I data. The comparison is shown
in four different bins in $Q^{2}=3.5,\,8.5,\,12$ and 18 GeV$^{2}$. 
In the DGLAP case the band corresponds to the PDF uncertainties.}
\label{all}
\end{figure}


\begin{figure}[t]
\begin{center}
\hspace{0.2cm}
\includegraphics[height=5.4cm]{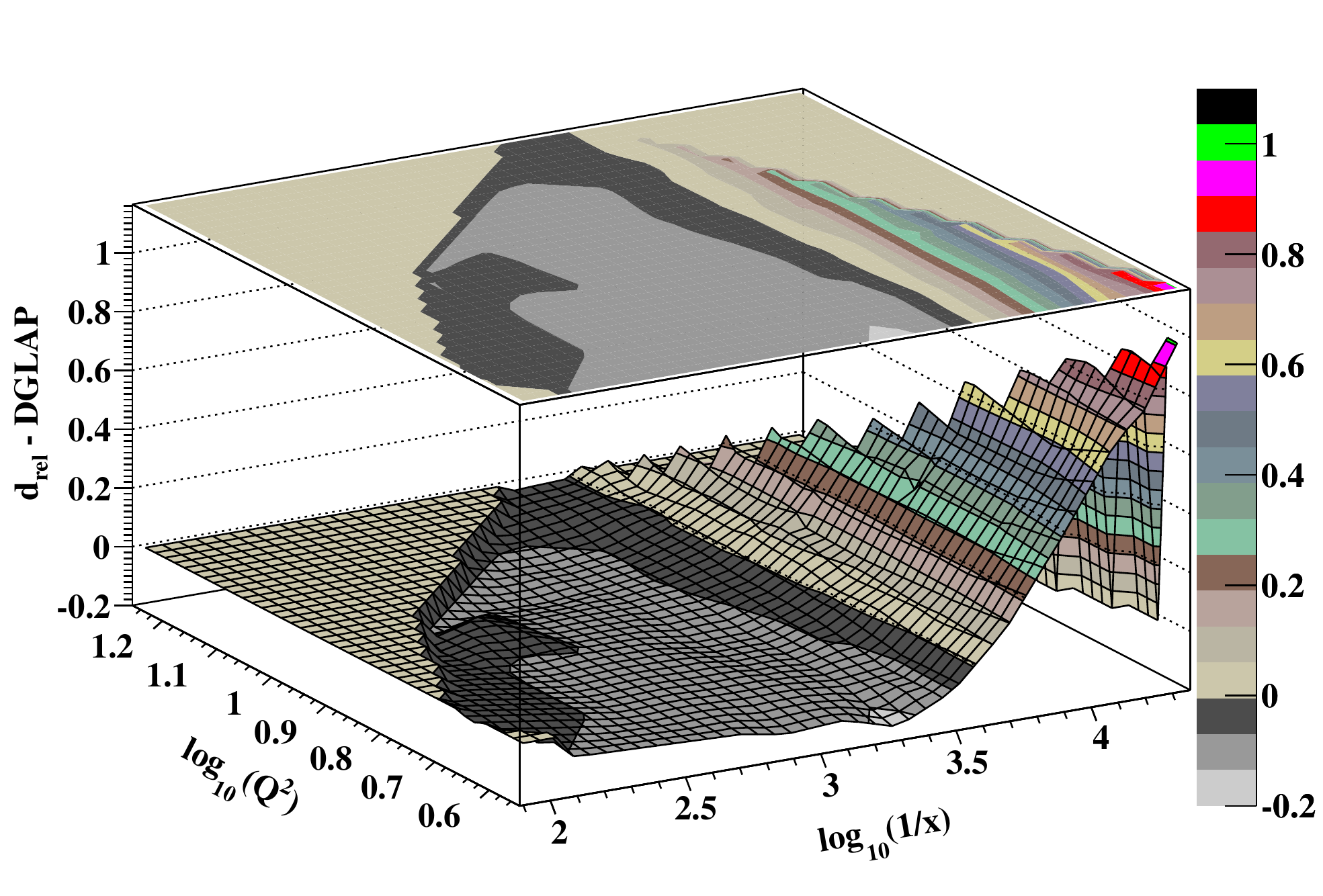}
\includegraphics[height=5.4cm]{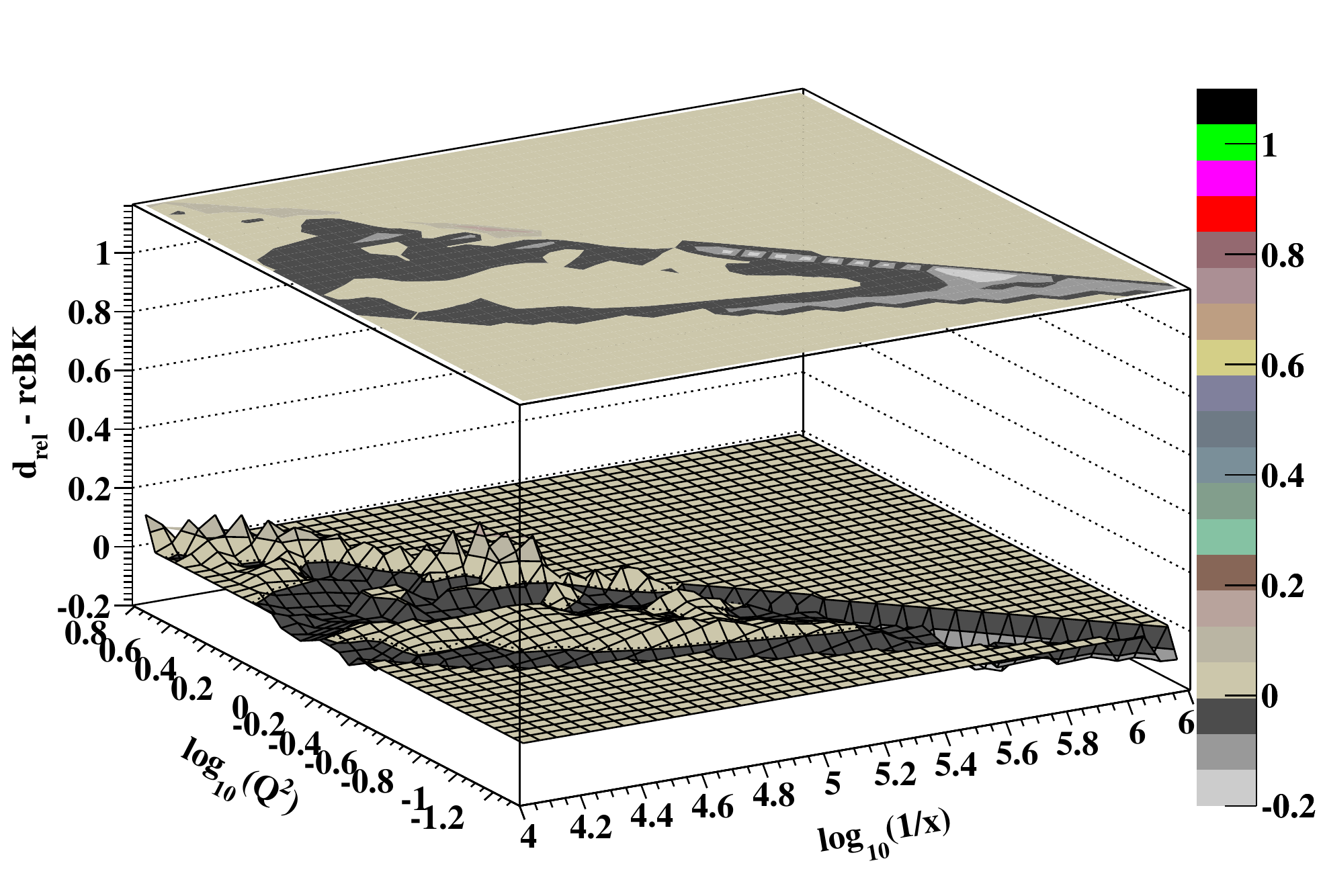}
\includegraphics[height=6.5cm]{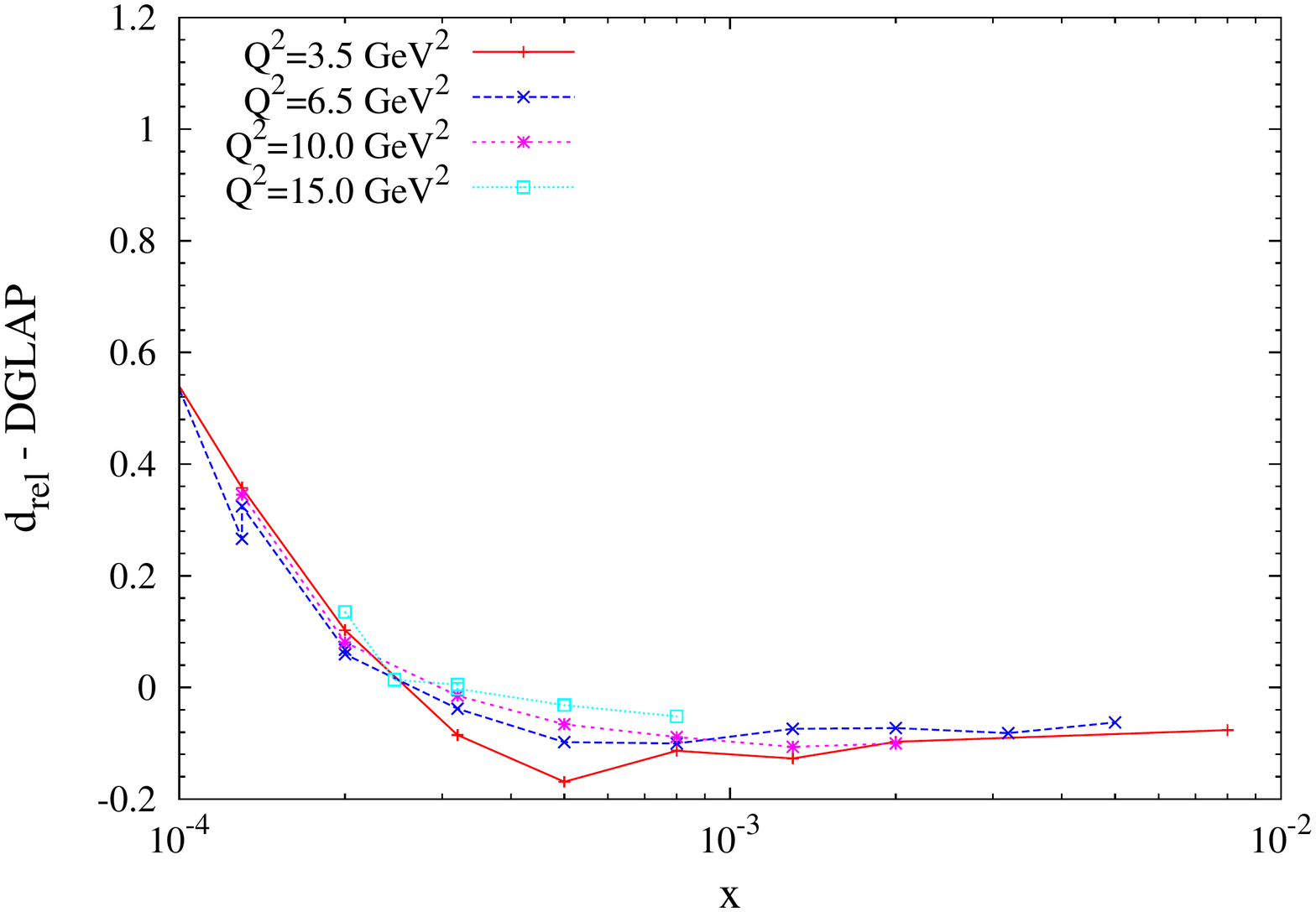}
\hspace{-1cm}
\includegraphics[height=6.5cm]{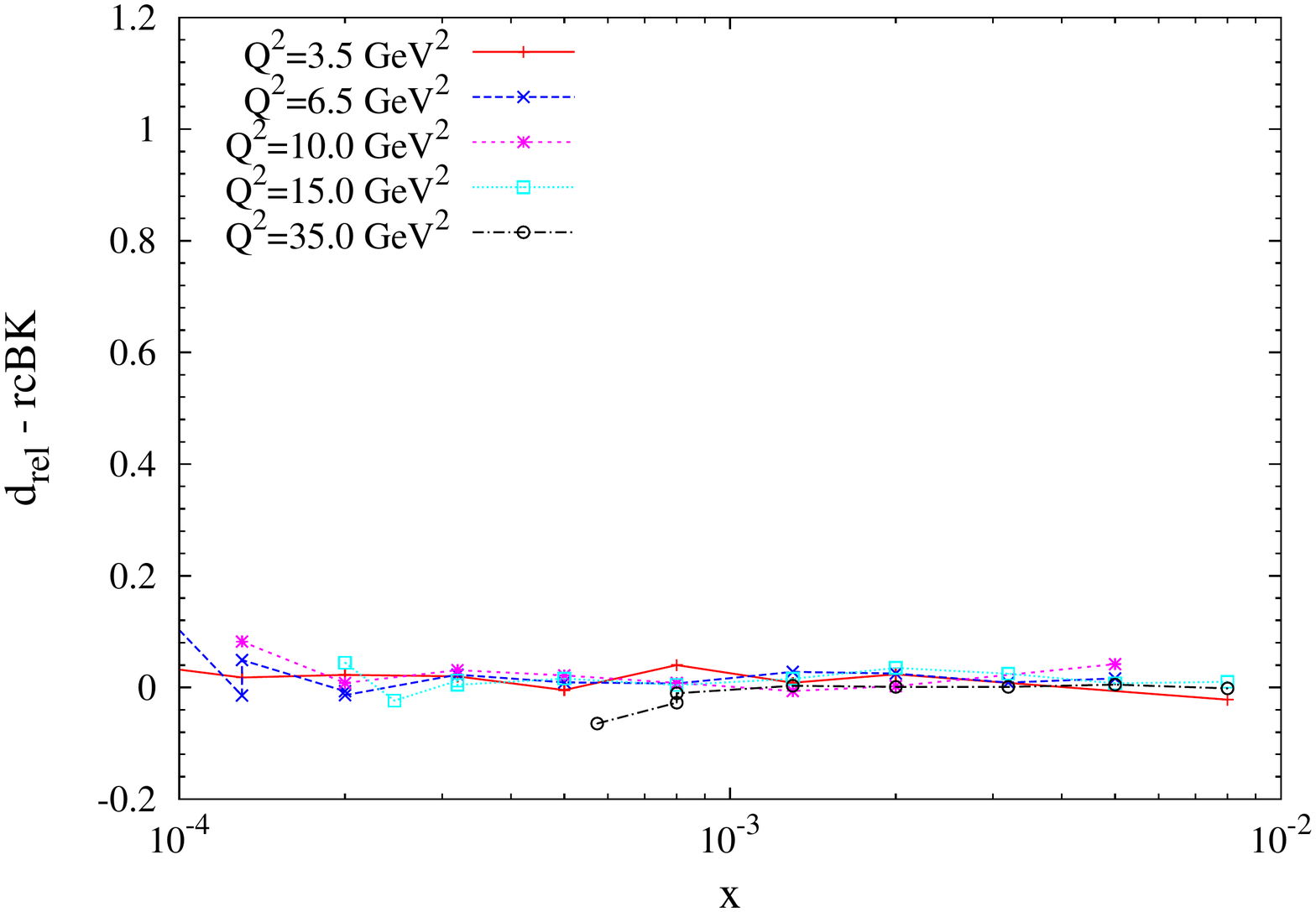}
\vspace{-1cm}
\end{center}
\caption{\small The relative distance, $d_{\rm rel}(x,Q^{2})$ Eq.~(\ref{eq:drel}), for DGLAP (left) and rcBK (right) cut fits.}
\label{drel}
\end{figure}

Fig.~\ref{all} shows our results corresponding to the rcBK fit with the most stringent cut $x_{\rm cut}=10^{-4}$ together with experimental data and the analogous results from the DGLAP fit with cut $A_{\rm cut}=1.5$. While the DGLAP extrapolations to the unfitted, {\it test} region are compatible with data within the uncertainty bands, the central values of the predictions show significant deviations from data in the region of small-$x$. Also, the PDF error bands blow up at small-$x$. This can be understood as a consequence of the fact that there is no data in the DGLAP fitting set causally connected to small values of $x$ or, equivalently, the DGLAP predictions at small values of $x$ are not empirically constrained by data. This explains the large error bars in this region.

In order to measure these deviations, we plot in Fig.~\ref{drel}
the relative distance between the theoretical results and experimental data both for the rcBK and DGLAP cut fits, the cut values being $x_{\rm cut}=10^{-4}$ and $A_{\rm cut}=1.5$ respectively. The relative distance is defined as
\begin{equation}
d_{\rm rel}(x,Q^{2})=\frac{\sigma_{\rm r,th}-\sigma_{\rm r,exp}}{(\sigma_{\rm r,th}+\sigma_{\rm r,exp})/2}\,,
\label{eq:drel}
\end{equation}
where $\sigma_{\rm r,th}$ are the theoretical predictions for the
HERA reduced cross sections and $\sigma_{\rm r,exp}$ the corresponding 
experimental data.
As shown in Fig.~\ref{drel},  $d_{\rm rel}(x,Q^{2})$ is on average much smaller for the rcBK fits than it is for the DGLAP one, the latter also showing a systematic trend to underestimate data at small-$x$ and to overshoot them at larger $x$. In turn, the rcBK values for $d_{\rm rel}(x,Q^{2})$ alternate in sign in all the unfitted region.


\begin{figure}[t]
\begin{center}
\hspace{0.2cm}
\includegraphics[height=5.4cm]{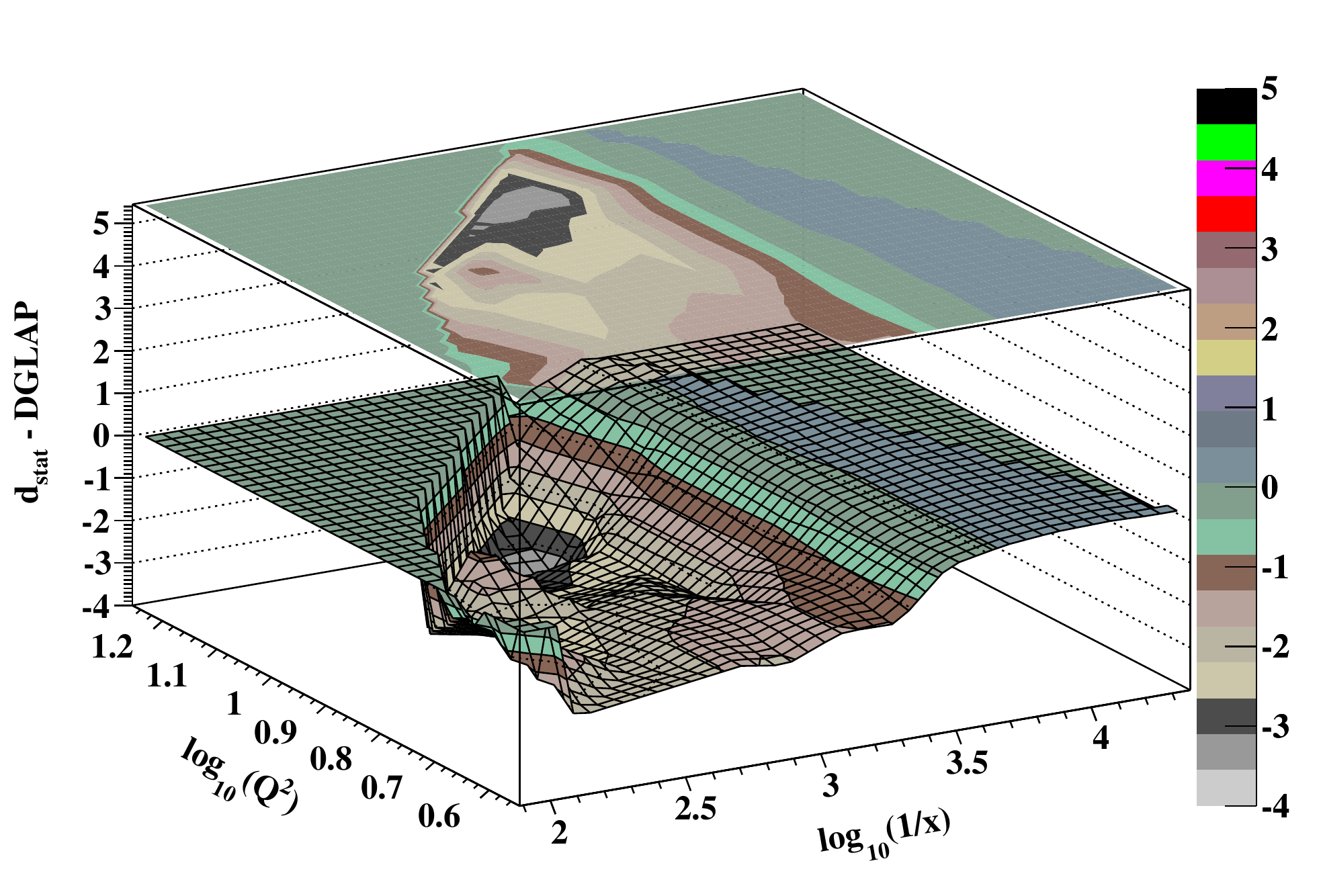}
\includegraphics[height=5.4cm]{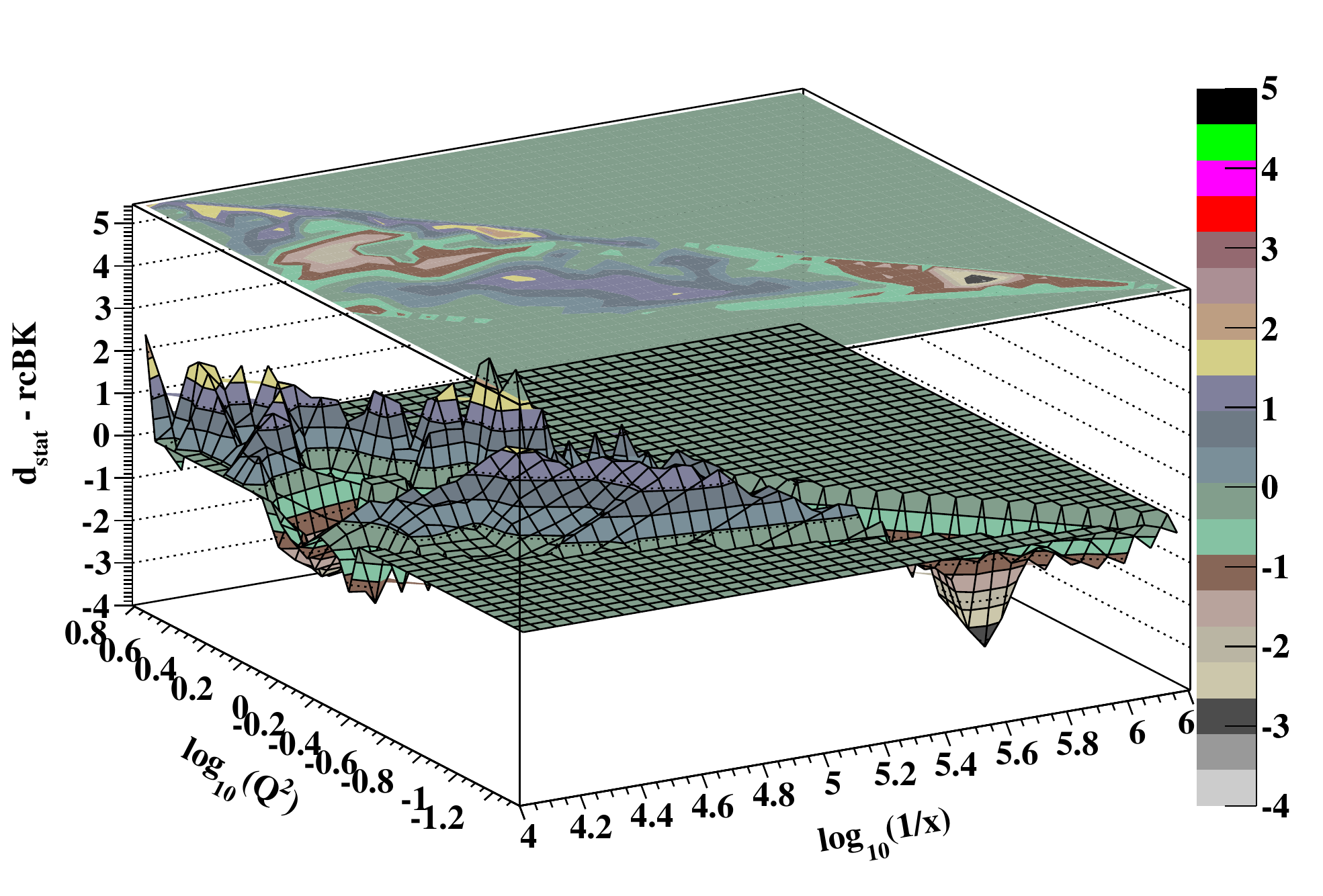}
\hspace{-2 cm} 
\includegraphics[height=6.5cm ]{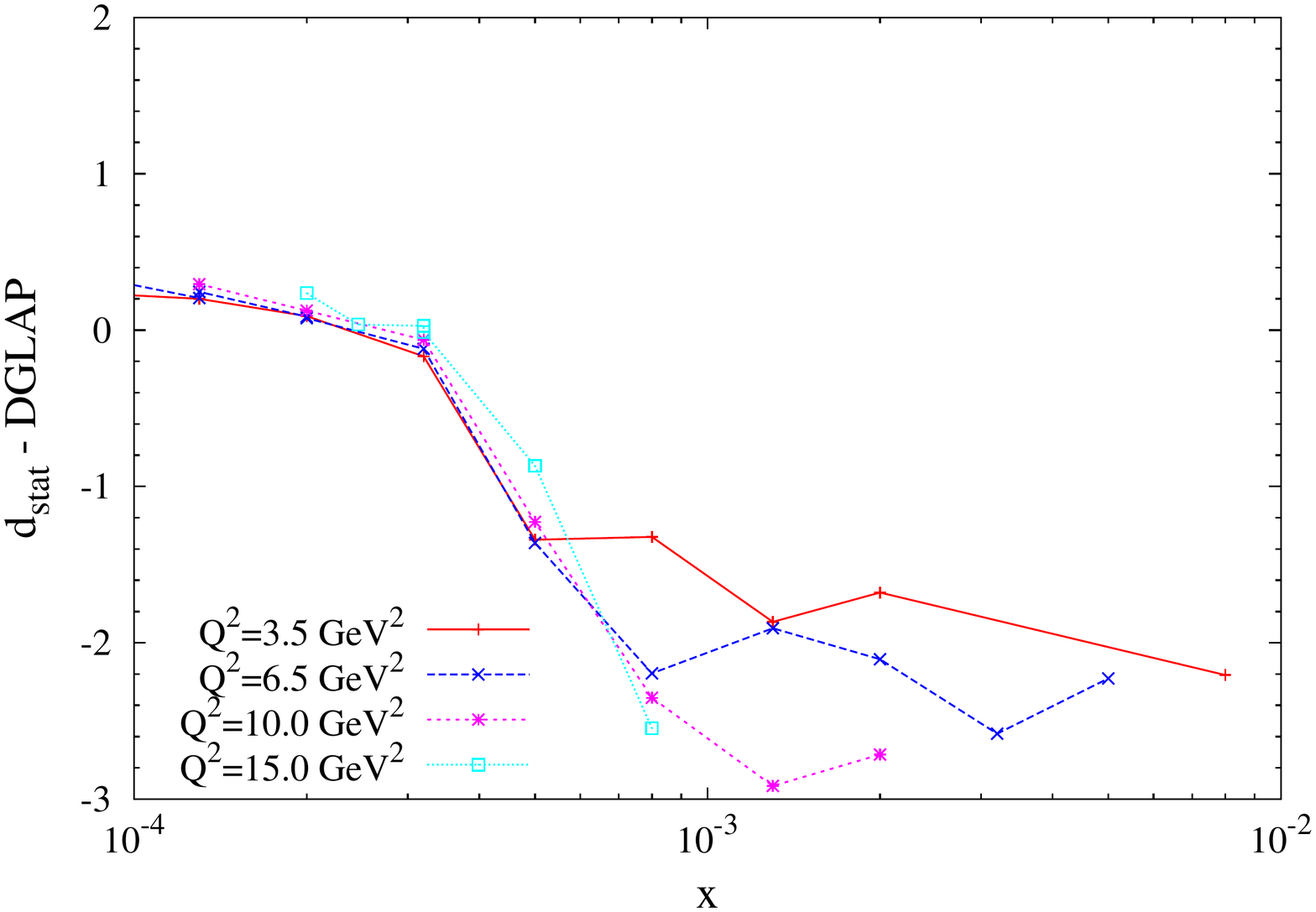}
\hspace{-1cm}
\includegraphics[height=6.5cm]{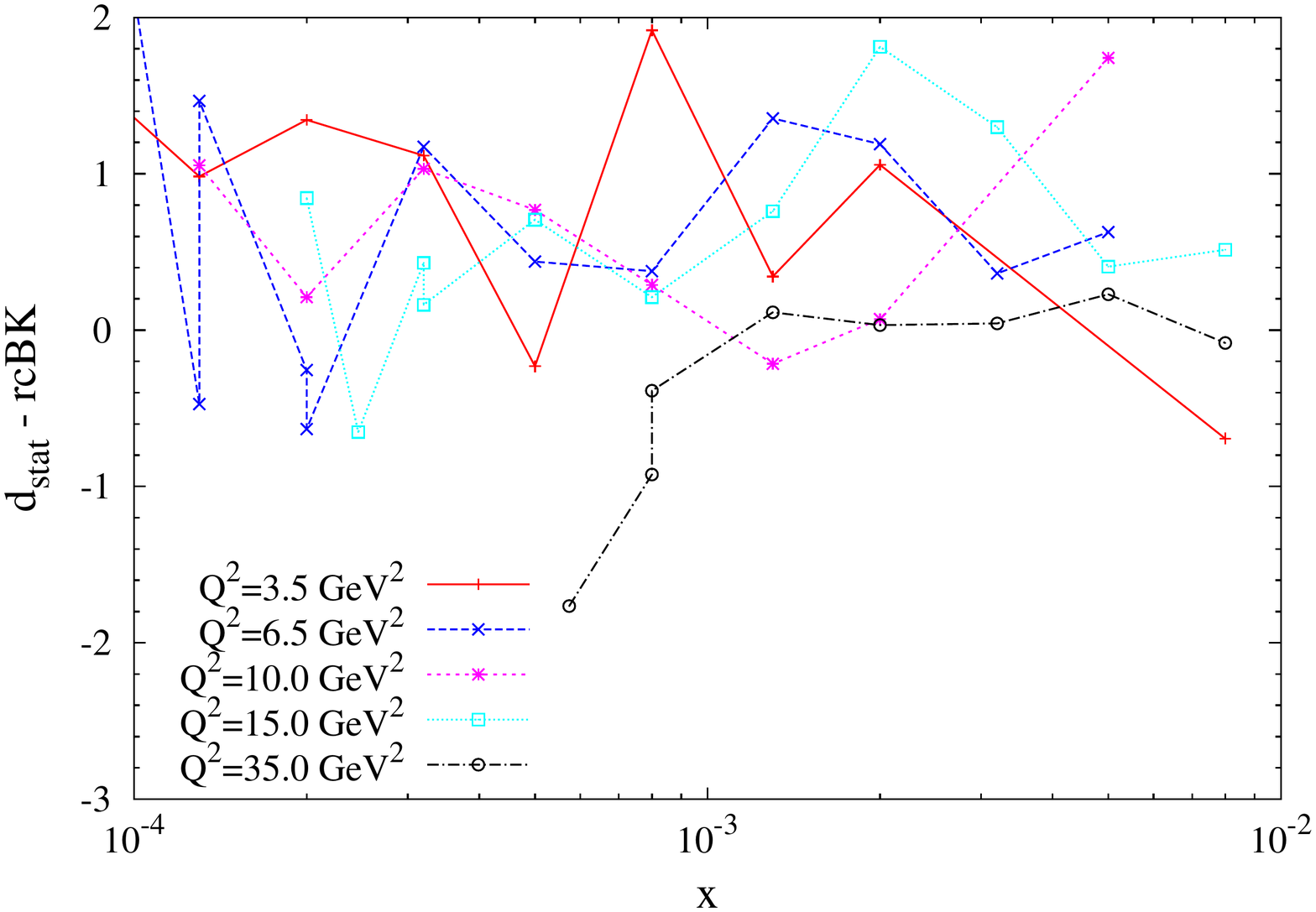}
\end{center}
\vspace{-1cm}
\caption{\small The statistical distance, $d_{\rm stat}(x,Q^{2})$  Eq.~(\ref{eq:dstat}), for DGLAP (left) and rcBK (right) cut fits. }
\label{dstat}
\end{figure}


 
\begin{figure}[h]
\begin{center}
\includegraphics[height=10cm]{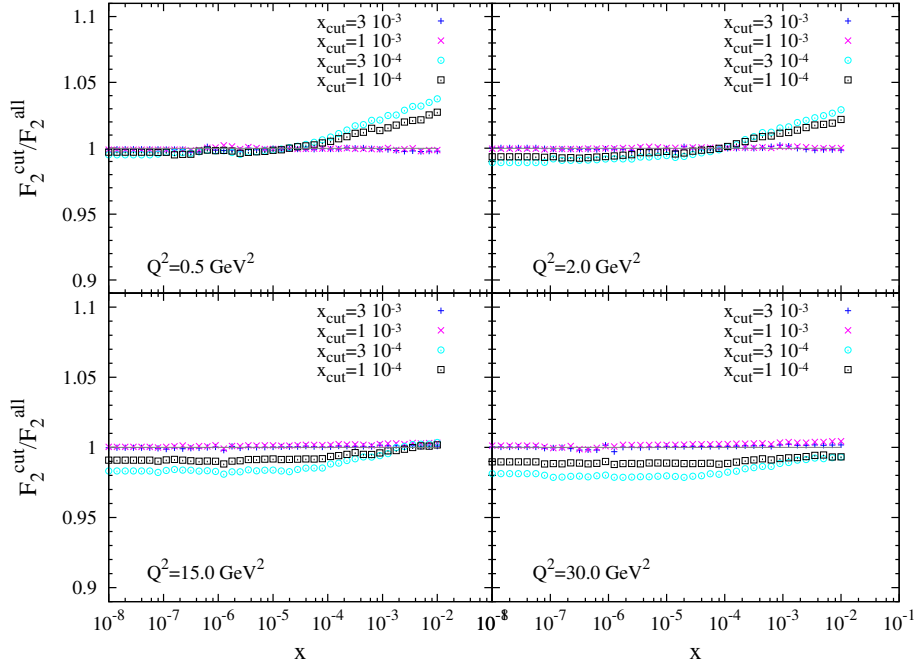}
\end{center}
\caption{\small Extrapolation to the low-x  region from the rcBK cut fits presented in Fig.~\ref{rcBKdata}. The total structure function, $F_2(x,Q^2)$,  is calculated down to $x=10^{-8}$. The results are presented as a ratio of the prediction for the different cut fits to the prediction for the uncut fit, i.e. a fit to all data with $x<x_0=10^{-2}$ and $Q^2<50$ GeV$^2$.}
\label{F2}
\end{figure}

\begin{figure}[h]
\begin{center}
\includegraphics[height=10cm]{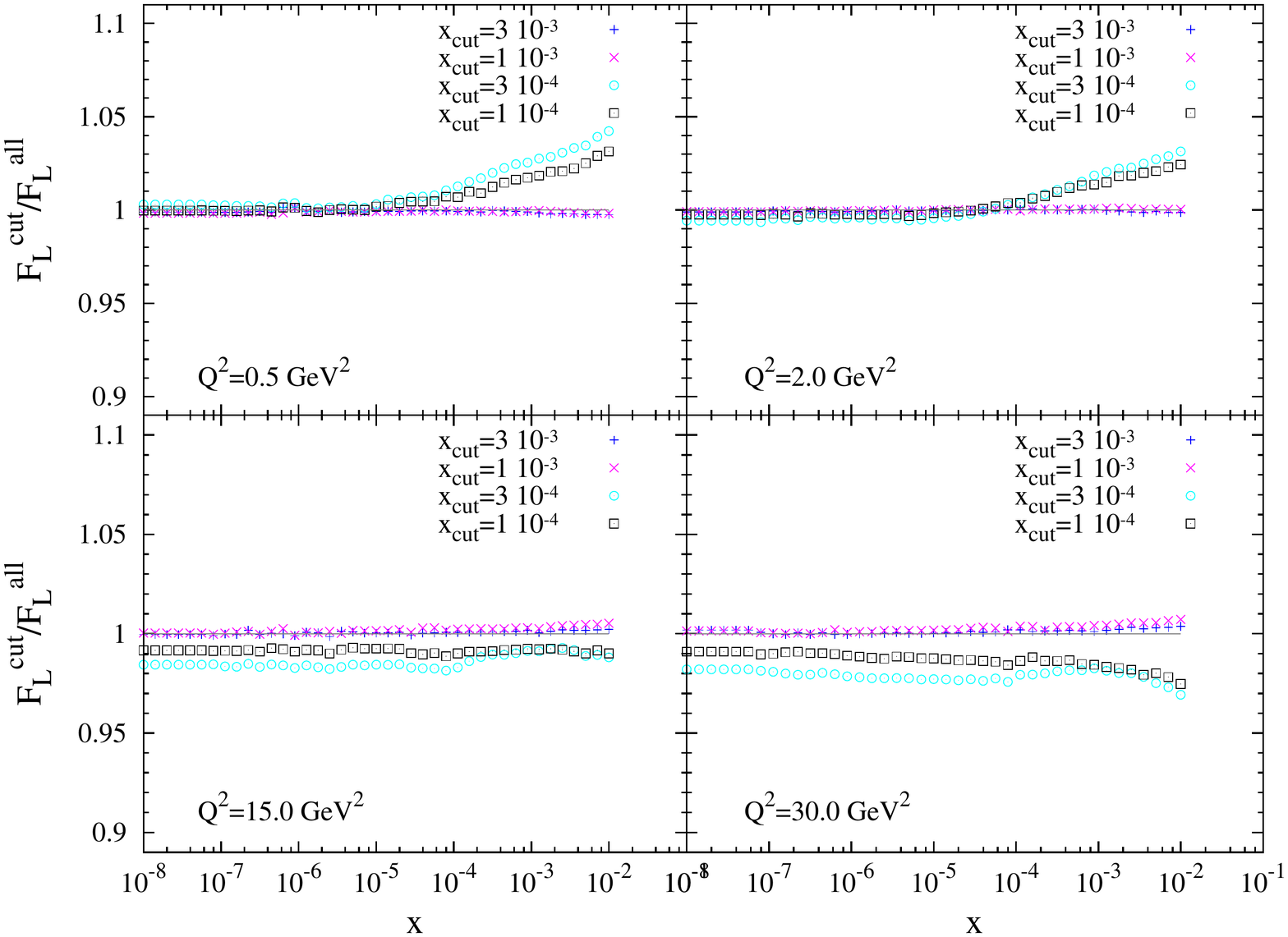}
\end{center}
\caption{\small Low-x extrapolation analogous to Fig.~\ref{F2} for the longitudinal structure function $F_L(x,Q^2)$.}
\label{FL}
\end{figure}

Measurements of the deviations of  theoretical predictions from data can be endowed with a more statistically meaningful measure, the statistical distance defined as
\begin{equation}
d_{\rm stat}(x,Q^{2})=\frac{\sigma_{\rm r,th}-\sigma_{\rm r,exp}}{\sqrt{\Delta\sigma_{\rm r,th}^{2}+\Delta\sigma_{\rm r,exp}^{2}}}\,.
\label{eq:dstat}
\end{equation}
While the relative distance Eq.~(\ref{eq:drel}) measures in absolute terms the deviation, Eq.~(\ref{eq:dstat}) gives the statistical significance of that deviation in units of the standard deviation.
The values of $d_{\rm stat}(x,Q^{2})$ for the rcBK fit and the DGLAP ones are shown in Fig.~\ref{dstat}.  
In the case of the rcBK fits the theoretical error has been estimated as the maximal difference among the theoretical predictions corresponding to fits with different cuts. 
The average distances for the rcBK fits with the most stringent cut, $x_{\rm cut}=10^{-4}$,  are
$\la d_{\rm rel}\ra= \lp 5  \pm 41\rp \cdot 10^{-3}$, and $\la d_{\rm stat}\ra= 0.3 \pm 9$. 
For the DGLAP cut fit, the average relative distance in the cut region
is $\la d_{\rm rel}\ra= 0.1 \pm 0.3$, while the statistical distance in the same region is  $\la d_{\rm stat}\ra=-0.8 \pm 1.1 $. 
Both $\langle d_{\rm rel}\rangle $ and $\langle d_{\rm stat} \rangle$ are considerably smaller for rcBK than they are for DGLAP, despite the fact that theoretical errors are probably underestimated in the rcBK approach. Note however that the initial conditions in the rcBK analysis are more restrictive than in the DGLAP
fit, and that adopting a more flexible input in the rcBK might affect the above
results.


In order to explore the predictive power of the rcBK approach and the sensitivity to boundary effects encoded in the different initial conditions for the evolution under the inclusion/exclusion of subsets of data we extrapolate our results for the total $F_{2}(x,Q^{2})$ and longitudinal  $F_{L}(x,Q^{2})$ structure functions to values of $x$ smaller than those currently available experimentally. The results of such extrapolation are presented in Figs.~\ref{F2} and~\ref{FL}. We find that the predictions stemming from different fits converge, within approximately one percent accuracy, at values of $x\sim10^{-4}$. The fit which extrapolation deviates the most, $\sim 2\div3 \%$, is the one corresponding to $x_{\rm cut}=3\cdot10^{-4}$, which also yields a larger $\chi^2/{\rm d.o.f.}$ to all data. This convergence is due to the fact that rcBK admit asymptotic solutions independent of initial conditions, as already known in the literature. 
Importantly, these predictions could be verified in planned facilities as the Large electron-Hadron Collider (LHeC)~\cite{lhec2} or the Electron Ion Collider~\cite{EICwhite}, where a much extended kinematic reach in $x$ would be available\footnote{In the context of future DIS facilities,
the physics that can be probed at low $x$ within the DGLAP framework has been studied in~\cite{Accardi:2011qh,Rojo:2009ut}.}.

\section{Implications for LHC phenomenology}
\label{implications}

Is follows from the previous discussion that if deviations from
fixed order DGLAP were conclusively found in HERA data, resulting from either the presence of
small-$x$ resummation or  non-linear effects, one should either
exclude this data from the DGLAP analysis or perform a new analysis
combining for instance fixed order DGLAP with small-$x$ resummation.\footnote{Of course such deviations from fixed order DGLAP might also contaminate extractions of the strong coupling constant from global PDF analyses, in 
which the HERA data play an important role~\cite{Ball:2011us}. }
To quantify the worst-case scenario,
in this final section we estimate 
the  theoretical
uncertainty stemming from these potential deviations in DGLAP fits. 
To do so we proceed along the lines of Refs.~\cite{Caola:2009iy,Caola:2010cy} and compute benchmark LHC cross sections with the  PDF sets both with and without the small--$x$ kinematical cuts, that is
using the same settings as in Ref.~\cite{Ball:2011uy}. 
We have computed, 
with the NNPDF2.1 NNLO set, 
the cross sections in NNLO QCD for electroweak gauge boson production
with {\tt VRAP}~\cite{Anastasiou:2003ds}, top quark production with 
{\tt HATHOR}~\cite{Aliev:2010zk} (based on an approximate NNLO calculation)
and Higgs production in 
gluon fusion with the code of Ref.~\cite{Bonciani:2007ex}. We consider
only PDF uncertainties, the strong coupling is kept fixed at its
reference value of
  $\alpha_s\lp M_Z\rp=0.119$ and the renormalization and factorization
scales are not varied.

\begin{table}[h]
\footnotesize
\centering
\begin{tabular}{|c|c|c|c|c|c|}
\hline
\multicolumn{6}{|c|}{LHC 7 TeV}  \\ 
\hline
\hline
 & $\sigma(W^+)B_{l\nu}$ (nb)
& $\sigma(W^-)B_{l\nu}$ (nb) 
& $\sigma(Z^0)B_{ll}$ (nb) 
& $\sigma(t\bar{t})$ (pb) 
& $\sigma(ggH)$ (pb) \\
\hline
NNPDF2.1 $A_{\rm cut}=0$ & $6.20 \pm 0.10$ & $4.21 \pm 0.07$  &$0.972 \pm 0.013$ & $ 167\pm 5$ &  $13.3  \pm 0.3 $ \\
NNPDF2.1  $A_{\rm cut}=1.5$ &  $6.13 \pm 0.17 $ & $4.17 \pm 0.10$  &$0.962 \pm 0.021$ & $ 171 \pm 7$ & $13.3  \pm 0.3 $\\
\hline
\end{tabular}

$\quad$

\begin{tabular}{|c|c|c|c|c|c|}
\hline
\multicolumn{6}{|c|}{LHC 14 TeV}  \\ 
\hline
\hline
 & $\sigma(W^+)B_{l\nu}$ (nb)
& $\sigma(W^-)B_{l\nu}$ (nb) 
& $\sigma(Z^0)B_{ll}$ (nb) 
& $\sigma(t\bar{t})$ (pb) 
& $\sigma(ggH)$ (pb) \\
\hline
NNPDF2.1 $A_{\rm cut}=0$& $ 12.45 \pm 0.22$ & $ 9.14 \pm 0.15$   & $ 2.08 \pm 0.03$ &  $ 935 \pm 17$ &  $44.1  \pm 0.5 $ \\
NNPDF2.1 $A_{\rm cut}=1.5$ & $12.69 \pm 1.19$ & $ 9.24 \pm 0.50$  &$2.10 \pm 0.11$ &  $920 \pm 30$ &  $44.0  \pm 0.7 $\\
\hline
\end{tabular}
\caption{\small Predictions for LHC cross sections, computed at NNLO QCD
with two different PDF sets: the 
reference NNPDF2.1 NNLO fit with $A_{\rm cut}=0$ and
with the NNPDF2.1 NNLO fit with $A_{\rm cut}=1.5$.
Upper table: LHC 7 TeV. Lower table: LHC 14 TeV. For
Higgs boson production a mass of $m_H=125$ GeV has been
used, while for top quark production we have used
 $m_t=172.5$ GeV. \label{tab:pheno}}
\end{table}

\begin{figure}[h!]
\centering
\epsfig{width=0.49\textwidth,figure=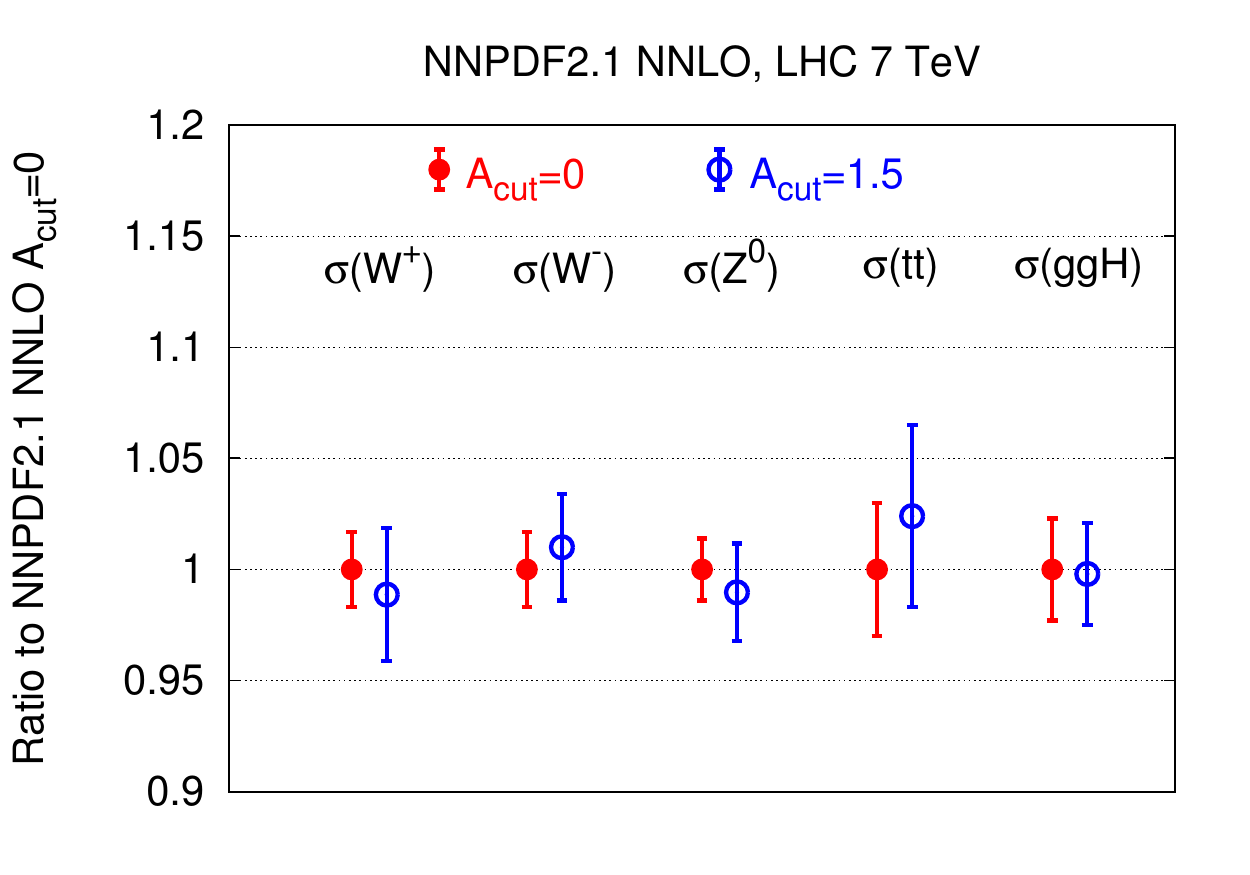}
\epsfig{width=0.49\textwidth,figure=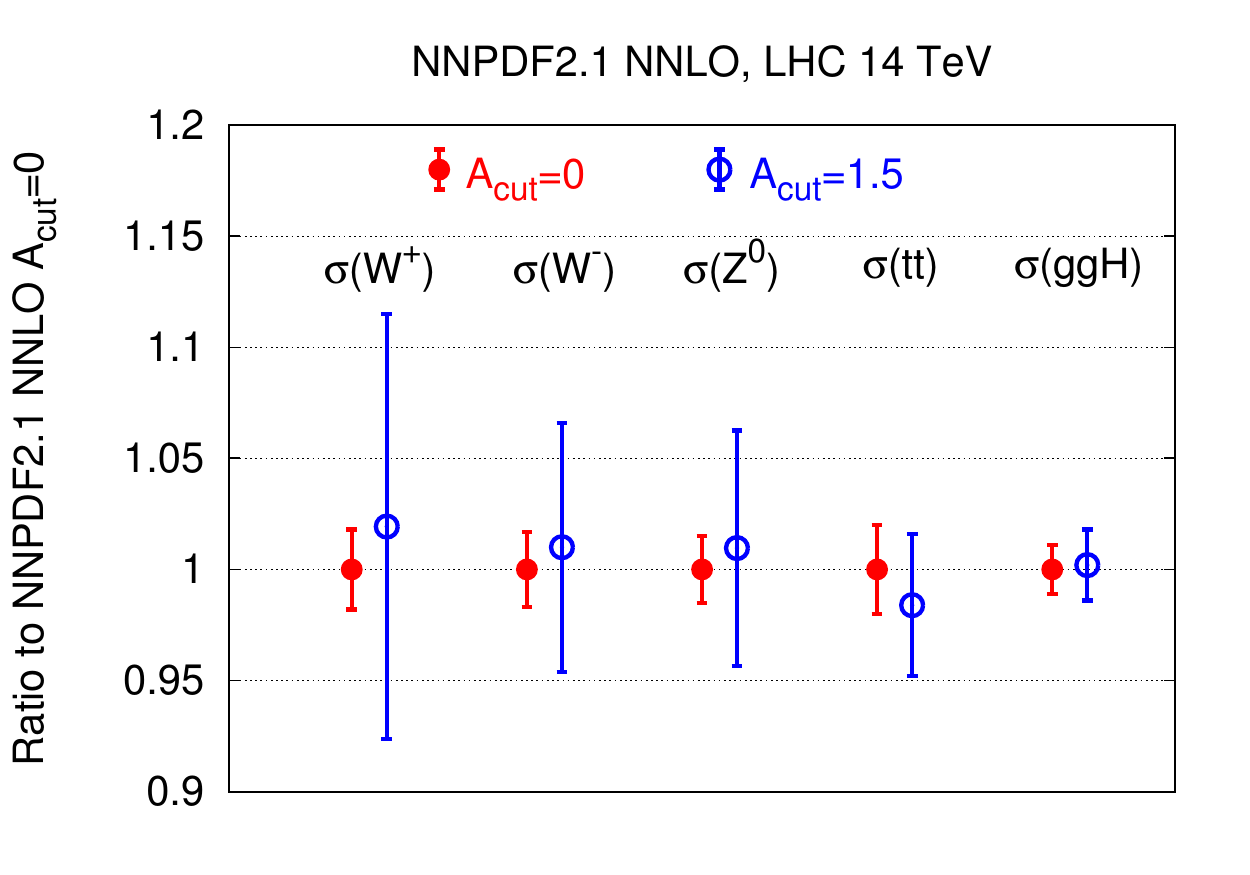}
\caption{\small Graphical summary of the results
of Table~\ref{tab:pheno}: 
comparison of the predictions for LHC NNLO cross sections
for the reference NNPDF2.1 NNLO fit with $A_{\rm cut}=0$ and
with the NNPDF2.1 NNLO fit with $A_{\rm cut}=1.5$. Cross sections are shown
as ratios to the uncut $A_{\rm cut}=0$ predictions. We show results
both for LHC 7 TeV (left plot) and for LHC 14 TeV (right plot).
\label{fig:pheno}}
\end{figure}

The results are summarized in Table~\ref{tab:pheno} and represented graphically
in Fig.~\ref{fig:pheno}. As can be seen, the impact of cutting the
small-$x$ and small-$Q^2$ HERA data from the fit is rather moderate at
LHC 7 TeV: 
the uncut results are always within the reference 1--sigma band and the
PDF uncertainties increase less than a factor two. On the other hand, 
the impact is much larger at LHC 14 TeV, since the larger center
of mass energy leads to smaller values of $x$ in the PDFs being probed, and 
these  are more affected  by the kinematical cut. Indeed,
for the electroweak boson production cross sections the PDF uncertainties
increase by up to a factor five, though the central predictions are still
in agreement with the original uncut results. On the other hand, the
cross section for Higgs boson production in gluon fusion is very
stable against the kinematical cuts.

These results suggest that understanding what is the correct dynamics
that drives the small-$x$ and $Q^2$ HERA data is important for
precision physics at the LHC specially once the machine energy
is raised up to its design value of 14 TeV.

\clearpage 

\section{Conclusions and outlook}\label{conclusions}

In summary, we have presented a precision study on the suitability of the rcBK and DGLAP approaches to describe HERA data in the kinematic region of moderate $x$ and $Q^{2}$.
Our strategy consists in setting a common {\it test} ground through selected kinematic cuts to the rcBK and DGLAP fitting procedures and perform systematic comparisons between theory extrapolations and the unfitted data set. Our main findings are:
\begin{itemize}
\item[i)] DGLAP fits display sensitivity to the exclusion of small-$x$ data from the fitted sample. While the deviations found are not statistically significant (below the 1--sigma level), this might give support to the general idea that novel interesting physics is being obscured by its encoding in the freedom of choice of initial conditions.
\item[ii)] rcBK fits are robust against the exclusion of data for values of $x$ down to $x\!=\!10^{-4}$ and for data with $Q^{2}\!<\!50$ GeV$^{2}$.
\item[iii)] The exclusion of small-$x$ data from DGLAP fits has consequences for LHC phenomenology, in particular in a significant increase on the theoretical uncertainty for standard production cross sections at a collision energy of 14 TeV.
\item[iv)] rcBK evolution yields robust predictions at small $x$ and thus can be decisively confronted with data from proposed facilities like the LHeC or the EIC. Within a DGLAP analysis uncertainties grow very fast for low $x$ outside the data region, and that renders the framework non predictive in that domain.
\end{itemize}



The results of this paper beg for the development of a framework in which both $Q^{2}$ (DGLAP) evolution and $x$ (rcBK) non-linear evolution are jointly accounted for, just as the common framework for the DGLAP and BFKL equations has been consistently obtained by the
small--$x$ resummation formalism. Ultimately, one would wish for a resumation of the non-linear kernel into DGLAP or alternatively to the encoding of $Q^{2}$ evolution into the B-JIMWLK framework. Unfortunately, at present no clear pathway in this direction exists.


At a more pragmatic level, the non-linear corrections implied by rcBK evolution could be encoded in an effective manner as a correction factor or further constrain to the results obtained within DGLAP.
For instance, the paucity of data at small-$x$ could be made up for by including rcBK generated pseudodata in DGLAP fits. Also, a thorough study of the impact of the small-$x$ resummed formalism in DGLAP fits would be very useful to delimit to what extent non-linear corrections embodied in rcBK evolution are actually necessary to describe present data.


\section*{Acknowledgments}
The research of J.L.A. is supported by a fellowship from the Th\'eorie LHC France initiative funded by the IN2P3. J.~R. is grateful to S.~Forte
for discussions. The research of J.~R.
has been supported by a Marie Curie Intra--European Fellowship
of the European Community's 7th Framework Programme under contract
number PIEF-GA-2010-272515. JGM acknowledge the support of Funda\c c\~ao para a Ci\^encia e a Tecnologia (Portugal) under project CERN/FP/116379/2010.  The work of PQA is funded by the French ANR under contract ANR-09-BLAN-0060.


\end{document}